\documentclass[sigconf,nonacm]{acmart}

\usepackage{amsmath}
\usepackage{graphicx}
\usepackage{subcaption}
\usepackage{tabularx}
\usepackage{multirow}
\usepackage{caption}
\usepackage[table]{xcolor}
\usepackage{makecell}
\usepackage[most]{tcolorbox}
\usepackage{booktabs}
\usepackage{hyperref}
\usepackage{array}
\usepackage{pifont}
\usepackage{wasysym}
\usepackage{soul}
\usepackage[normalem]{ulem}
\usepackage{pgfplots}
\usepackage{tikz}
\pgfplotsset{compat=1.18}

\AtBeginDocument{%
  }

\newcommand{\sys}{\textsc{MTD-Playground}}
\setcopyright{acmlicensed}
\copyrightyear{2018}
\acmYear{2018}
\acmDOI{XXXXXXX.XXXXXXX}
\acmConference[Conference acronym 'XX]{Make sure to enter the correct
  conference title from your rights confirmation email}{June 03--05,
  2018}{Woodstock, NY}
\acmISBN{978-1-4503-XXXX-X/2018/06}

\begin{document}

\tcbset{
  attackbox/.style={
    colframe=black,
    colback=gray!20,
    boxrule=0.7pt,
    boxsep=2pt,
    left=4pt,
    right=4pt,
    top=4pt,
    bottom=4pt
  }
}
\title{MTD-Playground: An Attacker-Aware Evaluation Framework for Network Moving Target Defense}
\author{
Mohammad Farhad$^{*\,\dagger}$,
Mohoshin Ara Tahera$^{*\,\dagger}$,
Padam Jung Thapa$^{*}$,
Shuvalaxmi Dass$^{*\,\dagger}$,
Bhupendra Acharya$^{*}$
}

\affiliation{
  \institution{*\ The Center for Advanced Computer Studies (CACS), School of Computing and Informatics,}
  \city{University of Louisiana at Lafayette}
  \state{Lafayette, LA 70504}
  \country{USA}
}

\affiliation{
  \institution{$\,\dagger$ AICSIL Research Lab, CACS, Lafayette, LA 70504, USA}
  \city{}
  \state{}
  \country{}
}

\email{
\{mohammad.farhad1, mohoshin-ara.tahera1, padam-jung.thapa1,\\shuvalaxmi.dass, 
bhupendra.acharya\} @ louisiana.edu
}

\renewcommand{\shortauthors}{M. Farhad et el.}

\begin{abstract}
Moving Target Defense (MTD) has emerged as a proactive network cyber defense paradigm that increases attacker uncertainty through dynamic network reconfiguration techniques such as Software-Defined Networking (SDN)-enabled path randomization. However, existing evaluations remain fragmented due to inconsistent attacker assumptions, attack scenarios, and evaluation metrics, limiting reproducibility and deployment-oriented comparison.
In this paper, we present \sys{}, an attacker-aware evaluation framework for benchmarking SDN-enabled path-randomization (PR) MTD techniques under small-scale realistic enterprise-style multi-stage attack scenarios. Beyond isolated security and performance metrics, \sys{} introduces a composite evaluation methodology for analyzing deployment effectiveness, mutation-interval trade-offs, and defender--attacker operational balance. Using periodic path randomization as a representative PR-MTD strategy, our evaluation shows that aggressive mutation intervals reduce attack success rates to 4--20\% while increasing attack completion time to 160--311\,s across evaluated attack scenarios. At the same time, PR-MTD improves throughput by up to 30.9\% and reduces internal-path latency without service interruption. Composite analysis further shows that shorter mutation intervals consistently achieve the highest deployment effectiveness and positive defender advantage. These results demonstrate that SDN-based PR-MTD can substantially disrupt multi-stage attack progression while remaining practically deployable in enterprise environments.
\end{abstract}





\maketitle

\section{Introduction}
\noindent Network security plays a critical role in protecting modern digital infrastructure, particularly modern enterprise networks that interconnect users, devices, applications, and data across distributed environments. However, these networks often expose predictable network behavior and static configurations fixed routing paths, static IP addressing, and long-lived communication patterns that can be exploited during multi-stage attack campaigns involving reconnaissance, lateral movement, distributed denial-of-service (DDoS), and data exfiltration attacks \cite{cisco,cloudflare_ddos}. Modern adversaries increasingly employ this multi-stage attack campaigns that exploit the static nature of traditional network infrastructures, where network paths, host identifiers, forwarding behavior, and service configurations often remain unchanged over time. Once attackers learn the structure and communication patterns of the network, this knowledge can be reused to facilitate subsequent attack stages with reduced uncertainty \cite{Jajodia2011,Ghaderi,Huang2020}. Traditional network defense mechanisms such as firewalls, intrusion detection and prevention systems (IDS/IPS), and static access-control policies primarily rely on static configurations and signature-based detection, which are often insufficient against adversaries in dynamic network environments \cite{Stein2018,Jafarian2015,Ahmed2016,Chai2020,Chowdhary2018}. To address these limitations,  Moving Target Defense (MTD) has emerged as a proactive network cyber defense paradigm that continuously changes network-visible elements, including network paths, IP addresses, topology views, and service locations, to invalidate attacker knowledge and increase adversarial uncertainty \cite{Gudla2020,Stein2018}. As a result, attackers must repeatedly restart reconnaissance and exploitation processes, increasing their operational cost and reducing the probability of successful compromise. Existing research has explored various MTD techniques, including network path randomization, IP mutation, and topology mutation, to increase attacker uncertainty and reduce attack progression \cite{Jafarian2015,Achleitner2016,Macwan2019}. These network-based MTD techniques are commonly characterized along three dimensions: \textit{what} network attributes are reconfigured (e.g., paths, IP addresses, ports, services, or virtual machines), \textit{how} reconfiguration is performed (e.g., mutation, randomization, migration, or deception), and \textit{when} reconfiguration occurs, including periodic, event-driven, or adaptive triggering strategies \cite{Carvalho2014,Sengupta2020,Zhuang2014}.\\
\indent Among enabling technologies for network-based MTD, Software-Defined Networking (SDN) is particularly well-suited for implementing dynamic MTD techniques due to its centralized control and programmable forwarding capabilities, which enable flexible traffic engineering and network path reconfiguration \cite{Hyder2021, Mayone2024}. SDN has increasingly been leveraged to enable MTD techniques such as IP mutation, topology mutation, virtual Machine (VM) migration and network path randomization where  IP mutation dynamically changes host identifiers to invalidate reconnaissance information \cite{Macwan2019,Gudla2020}, while topology mutation alters network views and connectivity patterns to disrupt attack planning \cite{Achleitner2016} VM migration relocates services or workloads to reduce attack persistence \cite{Ahmed2016}.\\ 
\indent Among SDN-enabled MTD techniques, network path randomization (PR) has emerged as one of the most practical and widely studied techniques due to its ability to dynamically modify forwarding network paths without requiring substantial changes to end hosts or applications \cite{Stein2018, Hyder2021, Chowdhary2017}. However, despite the growing research interest in path randomization, existing evaluations are often performed under heterogeneous experimental settings with inconsistent network topologies, attack scenarios, attacker assumptions, threat models and evaluation metrics \cite{Abdelkhalek2022, Fronczyk2025, Aydeger2025}. Such inconsistencies limit reproducibility and make fair comparison across proposed approaches difficult.\\
\indent To address this gap, we present \sys{}, a standardized attacker-aware evaluation framework designed to evaluate  SDN-enabled path-randomization MTD techniques under varying path mutation intervals measured in seconds (e.g., 10s, 20s, and 30s) within a controlled and reproducible experimental environment. The framework models a realistic enterprise network architecture consisting of an external attacker, perimeter firewall, public-facing DMZ web server, internal application server, database server, and internal client workstations. Within this environment, attackers execute multi-stage intrusion campaigns involving reconnaissance, initial compromise, lateral movement, privilege escalation, and data exfiltration. \sys{} provides standardized attack workflows and evaluation settings for fair and controlled assessment of SDN-enabled PR MTD techniques under consistent attacker and network conditions. This design enables systematic analysis of defensive effectiveness and security--performance trade-offs in realistic enterprise-scale network environments.\\

\noindent \textbf{Contributions.} In summary, this paper makes the following contributions:
\begin{itemize}
    \item We design \sys{}, an attacker-aware framework for benchmarking SDN-enabled path-randomization (PR) MTD techniques under realistic enterprise-style multi-stage attack scenarios.
    \item We model three representative attack workflows -- RCE-driven, SSRF-driven, and credential-based compromise -- to analyze how reconnaissance dependence influences PR-MTD effectiveness during multi-stage attack progression.
    \item We provide a controlled and reproducible SDN-based environment for systematic evaluation of PR-MTD techniques under standardized security and performance conditions.
    \item We develop a composite evaluation methodology that jointly analyzes deployment effectiveness, mutation-interval trade-offs, and defender--attacker operational balance through unified deployment-oriented metrics.
    \item Using periodic path randomization as a representative PR-MTD strategy, we show that aggressive mutation intervals reduce attack success rates to 4--20\% and increase attack completion time to 160--311\,s while achieving 2.1--3.8$\times$ higher security benefit than operational cost. Our results further show that SDN-based PR-MTD can improve throughput and reduce internal-path latency without service interruption.

\end{itemize}

\noindent \textbf{Scope.}
This study is conducted in a small-scale enterprise network and evaluates SDN-enabled path-randomization MTD, as it is one of the most widely studied and well-understood SDN-based MTD techniques \cite{Gudla2020,Narantuya2019,Fronczyk2025,Chowdhary2018}, making it a suitable starting point for establishing our evaluation methodology and validating \sys{} before extending to other mechanisms. Although the current implementation focuses on path randomization, \sys{} provides an extensible implementation that users can modify to develop and evaluate other MTD mechanisms, such as IP or port mutation, using the same security, performance, and deployment-effectiveness methodology.\\


Section \ref{sec:background} provides background and related work. Section \ref{sec:playground} describes the MTD-Playground architecture, attack scenarios, and threat model. Section~\ref{sec:evaluation} presents the evaluation methodology, including existing metrics and the composite scoring formulation. Section \ref{sec:eval} reports the security, performance, and composite results. Section \ref{sec:discussion} discusses findings, explains key observations. Section \ref{sec:limit} identifies limitations and future work. Section \ref{sec:conclusion} concludes the paper.

\section{Background \& Related Work}
\label{sec:background}
\subsection{Background}

In network security, SDN provides a programmable platform for implementing network MTD techniques through centralized control and dynamic forwarding. This capability is achieved by separating the \textit{control plane}, which manages routing decisions and network policies, from the \textit{data plane}, which handles packet forwarding based on controller-installed rules. As a result, SDN enables controllers to dynamically update routing policies, forwarding rules, and network reachability relationships without manual device reconfiguration \cite{McKeown2008}. This flexibility enables rapid adaptation of network behavior under evolving operational and security conditions \cite{Achleitner2016,Gudla2020}. By dynamically modifying forwarding behavior and network visibility, these techniques increase attacker uncertainty and reduce network predictability during reconnaissance and lateral movement \cite{Ankur16}.\\
\indent Prior work has explored diverse network-level MTD techniques, particularly SDN-enabled path-randomization (PR) mechanisms, to increase attacker uncertainty and reduce network predictability during reconnaissance and lateral movement \cite{Achleitner2016,Ankur16}. Existing PR techniques differ in both \textit{how} and \textit{when} path mutations are performed, including periodic mutation at fixed intervals \cite{Narantuya2019}, adaptive mutation based on observed conditions \cite{Gudla2020}, and trigger-based mutation in response to events such as detected scanning activity \cite{Achleitner2016,Hyder2021}. This work specifically evaluates \textit{fixed-interval} periodic path-randomization MTD techniques.\\
\indent \textbf{Limitations.} Existing PR-MTD evaluations remain fragmented and non-standardized due to heterogeneous experimental environments, inconsistent attacker assumptions, simplified attack scenarios, and incompatible evaluation methodologies. Prior studies further rely on diverse security and performance metrics under non-comparable conditions, making fair benchmarking difficult (Table~\ref{tab:metric_landscape}, Section~\ref{sec:related-work}). Consequently, the deployment-level effectiveness and operational trade-offs of PR-based MTD techniques under realistic multi-stage enterprise attacks remain insufficiently understood. Rather than proposing a new MTD mechanism, our goal is to provide a controlled, attacker-aware, and reproducible framework for systematically benchmarking PR-based MTD techniques under standardized conditions.

\subsection{Related Work}
\label{sec:related-work}

\noindent\textbf{Evaluation Environments.} Studies use different SDN controllers (POX \cite{Jafarian2015,Achleitner2016}, Ryu \cite{Macwan2019,Gudla2020}, ONOS \cite{Narantuya2019,Hyder2021,Abdelkhalek2022}, OpenDaylight \cite{Chowdhary2018,Chowdhary2017}), different emulation platforms (Mininet \cite{Jafarian2015,Macwan2019,Gudla2020}, Containernet \cite{Fronczyk2025}, custom testbeds \cite{Stein2018,Ahmed2016}), and different topology sizes (ranging from 3 hosts to multi-ISP networks). No two studies run on the same infrastructure, making it impossible to attribute performance differences to the MTD mechanism rather than the test environment. MTD-Playground uses a standardised Containernet testbed with ONOS controller that any researcher can reproduce.

\noindent\textbf{Threat Models.} Most studies evaluate against a single attack type, typically reconnaissance scanning \cite{Jafarian2015,Macwan2019,Gudla2020} or DDoS \cite{Stein2018,Hyder2021,Chowdhary2017,Chai2020}. Only two studies \cite{Chowdhary2018,Achleitner2016} model multi-stage attack progression, and none evaluate against multiple distinct attack techniques (RCE, SSRF, credential-based) under the same MTD mechanism. Attacker knowledge assumptions range from fully external \cite{Jafarian2015} to partial insider \cite{Achleitner2016}, but these differences are rarely made explicit. MTD-Playground evaluates three distinct multi-stage attack scenarios with clearly defined attacker models.

\noindent\textbf{Evaluation Metrics.} As detailed in Section \ref{sec:existing_metrics} and Table \ref{tab:metric_landscape}, the metrics used across studies are inconsistent and incomplete. Reconnaissance accuracy is reported by around 70\% studies but each defines it differently --- deprecation ratio \cite{Jafarian2015}, host count \cite{Achleitner2016}, entropy \cite{Stein2018}, threat score~\cite{Chowdhary2018} --- preventing direct comparison. Throughput is quantified by only 10\% \cite{Aydeger2025}. Controller overhead is discussed by all but measured with actual CPU or memory numbers by none. Most critically, no study provides a composite score that combines security gain and operational cost into a single deployment decision. MTD-Playground measures all metrics under identical conditions and introduces the effectiveness ratio $E$ to fill this gap.\\
\noindent\textbf{Security--Performance Tradeoff.} 90\% studies acknowledge that MTD creates a tradeoff between security and performance, with descriptions ranging from ``rapid mutation increases security but decreases performance''~\cite{Gudla2020} to ``faster mutation improves security but increases retransmissions''~\cite{Fronczyk2025}. However, these observations remain qualitative, no study provides a composite metric to determine whether the security benefit justifies the operational cost. Our work addresses this by introducing deployment-oriented composite metrics that jointly capture security disruption and operational overhead (Section~\ref{sec:composite_formulation}).\\
\noindent\textbf{Benchmarking and Reproducibility.} A recurring limitation across all studies is the lack of reproducible benchmarking infrastructure. Each study builds its own evaluation environment from scratch, uses different tools and configurations, and reports results that cannot be independently verified or compared against other mechanisms. Code and data are rarely made publicly available.  MTD-Playground addresses this by providing an open, containerized testbed with standardized attack scenarios, threat model, and a composite evaluation framework that produces comparable scores across different MTD techniques. 
Appendix \ref{app:mtd_comparison}, provides a detailed comparison across representative studies.\\
\indent Beyond SDN-enabled path-randomization mechanisms, recent studies have explored game-theoretic (GT) and AI/ML-driven MTD approaches for randomization timing (\textit{when}) and network reconfiguration under dynamic attack conditions (\textit{how}). Game-theoretic approaches model strategic attacker--defender interactions using optimization, stochastic, and Markov-based formulations to capture adaptive adversarial behavior and multi-stage attack progression \cite{Prakash2015,optimal_timing_stackelberg,adaptive_markov_game,dynamic_markov_differential,markov_robust_game}. AI/ML-driven approaches further employ reinforcement learning and adaptive policy learning to optimize mutation strategies and reconfiguration policies \cite{Chai2020,Eghtesad2020,marl_markov_mtd,Zhou2025, dass1}, while hybrid methods combine strategic modeling with adaptive execution \cite{Li2020,Prakash2015,dass2}. Across these approaches, existing studies commonly rely on simulation-based or heterogeneous experimental environments with inconsistent attacker assumptions, limited enterprise-scale attack scenarios, and reduced reproducibility.

\section{MTD-Playground Design}
\label{sec:playground}
\noindent This section presents the overall design of \sys{} framework --- small-scale enterprise system design and threat model including multi-stage intrusion workflow, attack vectors and representative attack scenarios.
\subsection{System Design}
\label{sec:system-model}

\noindent 
The enterprise topology consists of an external attacker, a public-facing DMZ web server, an internal application server, a backend database server containing sensitive data, an internal client workstation, and four SDN-enabled OpenFlow switches ($S_1$--$S_4$). The switches provide centralized programmable forwarding control and enable dynamic path randomization through SDN-based flow-rule management. Although compact, the topology captures common enterprise deployment patterns and enables evaluation of representative intrusion workflows including reconnaissance, initial access, lateral movement, privilege escalation, and data exfiltration. The framework further provides a standardized threat model, attack scenarios and path randomization settings to support fair and reproducible assessment of SDN-enabled path-randomization MTD techniques under comparable adversarial conditions.

\subsection{Threat Model} 

\begin{figure*}[htbp]
    \centering
    \includegraphics[width=0.9\textwidth]{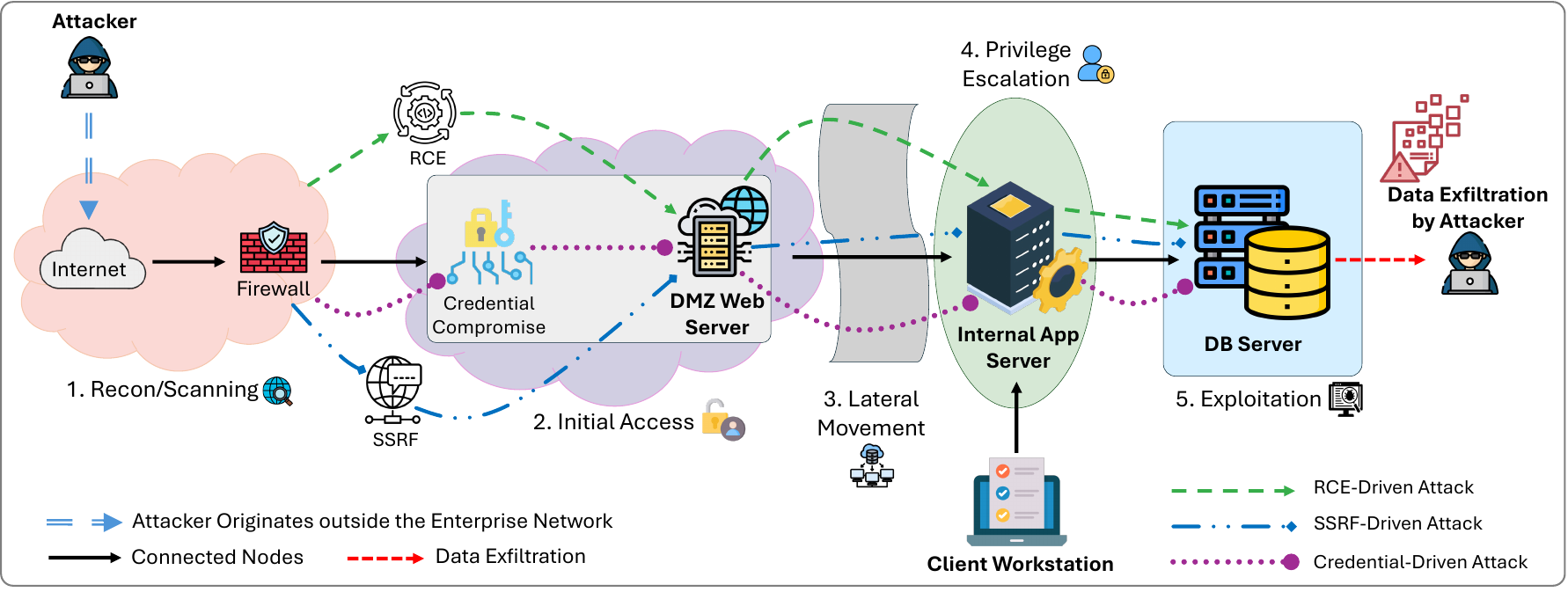} 
    \caption{Enterprise System Model and Multi-Stage Intrusion Workflow for different Attack Scenarios.} 
    \label{fig:system-model-multi-stage}
\end{figure*}

\noindent We consider a motivated external adversary operating against a realistic enterprise network environment. Unlike prior MTD studies that assume static or naive attackers, we model a limited-knowledge adversary. The adversary is aware that the defended network employs dynamic network defenses but lacks prior knowledge of the internal topology, current randomized network state, reconfiguration schedules, or defender-side control-plane telemetry. The attacker’s objective is to compromise the backend database and exfiltrate sensitive information by traversing a multi-tier enterprise architecture consisting of an Internet-facing DMZ web server, internal application server, and backend database server (details in Figure \ref{fig:system-model-multi-stage}). We further consider publicly exposed services on the DMZ web server to contain exploitable vulnerabilities that may be targeted during attacker reconnaissance and initial compromise attempts.\\
\indent The defender controls the SDN infrastructure, firewall policies, and PR-MTD logic, while monitoring standard operational telemetry such as flow records and host logs. Because perfect intrusion detection is not assumed, our evaluation focuses on how attackers progress across multiple intrusion stages rather than on isolated exploits, enabling assessment of how path-randomization MTD techniques disrupt adversarial movement within the enterprise network. To emulate realistic enterprise intrusions, we model representative attack vectors and attack scenarios spanning different stages of the intrusion workflow.

\subsubsection{Multi-stage Intrusion Workflow}
\label{sec: multi-stage-workflow}
\noindent
To evaluate the resilience of path-randomization MTD techniques, we model enterprise intrusion workflows involving reconnaissance, initial access, lateral movement, privilege escalation, and data exfiltration stages (Figure \ref{fig:system-model-multi-stage}).\\
\noindent \textbf{Reconnaissance.}
The attacker performs host discovery, service enumeration, and network scanning to identify reachable services and infer network structure.\\
\noindent \textbf{Initial Access.}
The attacker attempts to gain an initial foothold on the Internet-facing DMZ web server through exposed services or vulnerable applications.\\
\noindent \textbf{Lateral Movement.}
Following initial compromise, the attacker attempts to pivot toward internal resources, particularly the application server and backend systems.\\
\noindent \textbf{Privilege Escalation.}
The attacker seeks elevated privileges to expand access, maintain persistence, and facilitate further compromise of internal assets.\\
\noindent \textbf{Exploitation.} The attacker attempts database exploitation by targeting vulnerabilities in the backend database server

\noindent \textbf{Data Exfiltration.}
Accessing database, the attacker try to exfiltrate sensitive information from the backend database server, representing the final objective of the attack campaign.

\subsubsection{Attack Vectors}
\label{sec:attack-vectors}
To emulate realistic enterprise attack scenarios, the adversary employs representative attack vectors including Remote Code Execution (RCE) \cite{rce}, Server-Side Request Forgery (SSRF) \cite{ssrf}, credential-based compromise \cite{credential_attack}. RCE, SSRF, and credential compromise are modeled as intrusion-oriented attack vectors that may be leveraged during different intrusion stages depending on the targeted service, exposed attack surface, and attacker objective. An adversary may additionally apply distributed denial-of-service (DDoS) \cite{cloudflare_ddos} threat to disrupt firewall availability and increase operational pressure during intrusion attempts, however DDoS mitigation effectiveness are not considered within the scope of the current evaluation. 

\subsubsection{Attack Scenarios}
\label{sec:attack-scenarios}
We model three  attack scenarios that combine different attack vectors across multiple intrusion stages to emulate realistic adversarial progression through the enterprise network.

\begin{tcolorbox}[attackbox]
\textbf{Scenario 1 (RCE-driven attack):} Recon $\rightarrow$ Initial Access on Web (RCE) $\rightarrow$ Lateral Movement to App $\rightarrow$ Privilege Escalation (Credential Theft) $\rightarrow$ DB Access $\rightarrow$ Data Exfiltration
\end{tcolorbox}

\noindent
\textbf{Explanation:}
The adversary identifies a vulnerable Internet-facing web service and exploits an RCE vulnerability to obtain an initial foothold on the DMZ web server. The attacker subsequently pivots to the internal application server, escalates privileges through credential theft or abuse, and ultimately accesses the backend database to exfiltrate sensitive data.

\begin{tcolorbox}[attackbox]
\textbf{Scenario 2 (SSRF-driven attack):}
Recon $\rightarrow$ Initial Access on Web $\rightarrow$ SSRF-based Reachability to App $\rightarrow$ App Compromise or Credential Abuse $\rightarrow$ DB Access $\rightarrow$ Data Exfiltration
\end{tcolorbox}

\noindent
\textbf{Explanation:}
The adversary exploits a vulnerable web application feature to trigger SSRF, enabling indirect access to internal services through trusted server-side communication. The attacker then compromises the internal application server or abuses exposed credentials to progress toward the backend database.

\begin{tcolorbox}[attackbox]
\textbf{Scenario 3 (Credential-driven attack):}
Recon $\rightarrow$ Credential Compromise $\rightarrow$ Privileged Web Access $\rightarrow$ Lateral Movement to App $\rightarrow$ Privilege Escalation $\rightarrow$ DB Access $\rightarrow$ Data Exfiltration
\end{tcolorbox}

\noindent
\textbf{Explanation:}
The adversary acquires valid credentials through phishing, credential leakage, or credential reuse to gain privileged access to the DMZ web server. Using this trusted access, the attacker performs lateral movement toward internal services, escalates privileges as needed, and ultimately accesses the backend database for data exfiltration.

\section{Evaluation}
\label{sec:evaluation}
\subsection{Experimental Setup}
\label{sec: experimental-setup}
\noindent \sys{} is implemented on a virtualized SDN testbed integrating network emulation, container-based services, and centralized SDN control to support repeatable evaluation of path-randomization MTD techniques.
\subsubsection{Topology Design}
The experimental environment is implemented using Mininet \cite{mininet2010} within a virtualized infrastructure to emulate an enterprise-style network topology. To support realistic service deployment, Containernet \cite{containernet} is used to integrate Docker \cite{docker} containers as emulated network hosts while preserving Mininet’s network emulation capabilities. The SDN infrastructure employs Open vSwitch (OVS) \cite{openvswitch} as the data-plane switch and the Open Network Operating System (ONOS) \cite{onos2014} as the centralized SDN controller responsible for flow-rule installation, network management, and dynamic enforcement of path-randomization policies.
The topology consists of four OpenFlow (OF13)-enabled switches, five host nodes, ten network links, and 58 flow rules supporting small scale enterprise-style communication and forwarding policies. Docker-based hosts emulate the attacker, DMZ web server, internal application server, and backend database server, while a standard Mininet host generates legitimate client activity. The testbed remains configurable, enabling evaluation of different mutation intervals, attack scenarios, and network configurations under consistent experimental conditions.
\subsubsection{Security Components}
\noindent
An ONOS-based Layer-2 (L2) firewall is deployed to enforce network access-control policies by filtering Ethernet frames based on MAC addresses, switch ports, and traffic direction. This mechanism restricts unauthorized communication between external and internal network segments unless explicitly permitted, reflecting common enterprise security practices. The firewall is centrally managed through the SDN controller, enabling dynamic policy updates during experimentation. Integrating the firewall within the SDN-controlled environment allows the framework to emulate realistic security enforcement while evaluating MTD behavior under dynamic network reconfiguration.

\noindent
\subsubsection{Attack Scenario Execution}

\noindent
All attack scenarios were executed end-to-end on the \sys{} testbed using real network traffic between containerized hosts. Each attack stage is carried out by interacting with deployed services and exploiting corresponding vulnerabilities (RCE, SSRF, credential-based compromise) rather than through abstract or scripted stage transitions. Progression to each subsequent stage requires successful completion of the preceding attack stage. Each scenario follows the same enterprise attack path (Attacker$\rightarrow$Web$\rightarrow$Internal App$\rightarrow$Database) and experiences identical PR-MTD reconfiguration behavior. The primary difference across scenarios lies in the number of internal transitions requiring active reconnaissance and exploitation, which directly influences mutation sensitivity and attacker disruption.

\subsubsection{Periodic MTD Configurations}
To evaluate the security--perf. trade-offs of SDN-enabled PR MTD techniques, we instantiate \sys{} using periodic path-randomization (PR) MTDs under multiple mutation intervals---10, 20, and 30 seconds -- chosen as a starting baseline consistent with commonly evaluated configurations in related work \cite{Stein2018,Jafarian2015,Moghaddam2024,Aydeger2025,Macwan2019}. The SDN controller dynamically updates forwarding paths and flow rules across OpenFlow switches to create time-varying network reachability while maintaining transparent communication for legitimate services.

These three mutation intervals were chosen for evaluation as they represent aggressive, moderate, and conservative PR-MTD configurations under identical attack conditions. Shorter mutation intervals increase attacker disruption and reduce reconnaissance consistency but introduce more frequent flow-rule updates and controller overhead. In contrast, longer mutation intervals improve operational stability at the cost of reduced network unpredictability. The configurations are represented as follows:

\begin{itemize}
    \item \noindent \textbf{MTD10s.}
  Forwarding paths are re-randomized every 10 seconds, representing an aggressive mutation strategy designed to maximize attacker disruption and minimize reconnaissance consistency.
    \item \noindent \textbf{MTD20s.}
  Forwarding paths are re-randomized every 20 seconds, representing a moderate mutation strategy that balances security effectiveness with operational stability.

    \item \noindent \textbf{MTD30s.}
     Forwarding paths are re-randomized every 30 seconds, representing a conservative mutation strategy that minimizes operational overhead while providing lower levels of path unpredictability. \\
\end{itemize}

\noindent \textbf{No-MTD.} No path randomization is applied (no MTD defense in place), and the network operates with static forwarding paths throughout the experiment. This configuration serves as the baseline for evaluating the overall performance of the proposed path-randomization strategies under identical attack scenarios.

\textit{Note:} At each mutation interval, the defender randomly selects one of the available network paths between communicating hosts. Path selection is independent of latency, throughput, or hop-count optimization, as the objective is to increase network unpredictability rather than optimize routing performance.


\subsection{Composite Evaluation Framework}
\label{sec:setup_composite}
\subsubsection{Existing Evaluation Metrics}
\label{sec:existing_metrics}

Prior MTD studies evaluate effectiveness using a diverse set of security and performance metrics. We organize these metrics into two categories: (i) security metrics that measure attacker disruption and defensive effectiveness, and (ii) performance metrics that capture operational overhead and network stability. Table~\ref{tab:metric_landscape} summarizes the evaluation metrics reported across representative MTD studies and distinguishes metrics quantitatively evaluated ($ \dagger $) from those only qualitatively discussed ($\sim$).

\newcommand{\quant}{\ensuremath{\dagger}}
\newcommand{\qual}{\ensuremath{\sim}}
\newcommand{\mtdyes}{\ensuremath{\checkmark}}
\newcommand{\mtdno}{--}

\begin{table}[!t]
\centering
\caption{Evaluation metrics across 15 representative MTD studies.
\quant\,= quantitative evaluation in prior work;
\qual\,= qualitative discussion only;
\mtdyes\,= evaluated by \sys{};
\mtdno\,= not evaluated.}
\label{tab:metric_landscape}
\footnotesize
\setlength{\tabcolsep}{3pt}
\renewcommand{\arraystretch}{1.15}

\begin{tabularx}{\columnwidth}{@{}p{0.6cm} X c@{}}
\hline
\textbf{Metric} &
\textbf{Reported by} &
\textbf{MTD-PLAYGROUND} \\
\hline

\multicolumn{3}{@{}l}
{\cellcolor{gray!15}\textit{\textbf{Security Metrics}}} \\
\hline

ASR &
\quant\,\cite{Jafarian2015,Achleitner2016,Narantuya2019,
Hyder2021,Chowdhary2017,Chai2020,Stein2018};
\qual\,\cite{Macwan2019,Gudla2020,Fronczyk2025,Ahmed2016,
Chowdhary2018,Abdelkhalek2022,Aydeger2025};
1 of 15 does not report
& \mtdyes \\

\rowcolor{gray!5}
ACT &
\quant\,\cite{Jafarian2015,Achleitner2016,Ahmed2016,Chai2020};
11 of 15 do not report
& \mtdyes \\

RA &
\quant\,\cite{Jafarian2015,Achleitner2016,Narantuya2019,
Hyder2021,Ahmed2016,Chowdhary2018,Chowdhary2017,
Stein2018,Fronczyk2025};
6 of 15 do not report; definitions vary across studies
& \mtdyes \\

\rowcolor{gray!5}
NU &
\quant\,\cite{Stein2018,Narantuya2019};
13 of 15 do not report
& \mtdyes \\

DT &
\quant\,\cite{Stein2018,Hyder2021,Achleitner2016,Moghaddam2024,Aydeger2025,
Chowdhary2017,Abdelkhalek2022};
8 of 15 do not report
& n/a$^{*}$ \\

\rowcolor{gray!5}
DE &
\qual\,\cite{Achleitner2016,Macwan2019,Gudla2020,
Chai2020,Fronczyk2025};
10 of 15 do not report; none quantify
& \mtdyes \\

\hline
\multicolumn{3}{@{}l}
{\cellcolor{gray!15}\textit{\textbf{Performance Metrics}}} \\
\hline

SL &
\quant\,\cite{Achleitner2016,Hyder2021,Ahmed2016,
Abdelkhalek2022,Aydeger2025};
10 of 15 do not report
& \mtdyes \\

\rowcolor{gray!5}
TH &
\quant\,\cite{Fronczyk2025, Aydeger2025,Chowdhary2017};
12 of 15 do not report
& \mtdyes \\

PLR &
\quant\,\cite{Abdelkhalek2022,Aydeger2025,Chai2020,
Fronczyk2025};
11 of 15 do not report
& \mtdyes \\

\rowcolor{gray!5}
SA &
\quant\,\cite{Abdelkhalek2022,Aydeger2025,
Hyder2021,Chowdhary2017,Chai2020};
\qual\,\cite{Macwan2019,Gudla2020,Narantuya2019,
Stein2018,Chowdhary2018};
5 of 15 do not report
& \mtdyes \\

RT &
\quant\,\cite{Ahmed2016,Chowdhary2017,Aydeger2025,
Fronczyk2025};
11 of 15 do not report
& \mtdno \\

\rowcolor{gray!5}
CO &
\qual\,All 15 representative studies;
none report measured CPU or memory usage
& \mtdyes \\


\hline
\end{tabularx}

\par\vspace{0.2em}
\raggedright\scriptsize
{$^{*}$Path mutation is schedule-based, not detection-triggered.}
\end{table}

\noindent\textit{Security Metrics.}
Commonly reported security metrics include Attack Success Rate (ASR), which measures the fraction of attacks reaching their objective; Attack Completion Time (ACT), which captures how long a multi-stage intrusion requires; and Reconnaissance Accuracy (RA), which reflects whether attacker reconnaissance remains valid under network randomization. Additional metrics include network unpredictability (NU), detection time (DT), data exposure (DE) which measures the attacker’s ability to infer the deployed MTD behavior.
 
\noindent\textit{Performance Metrics.}
Performance evaluation commonly considers service latency (SL), throughput (TH), packet-loss rate (PLR), service availability (SA), recovery time (RT), and controller overhead (CO), including controller CPU, memory, and flow-rule management costs. Several studies also evaluate scalability and stability under dynamic network reconfiguration.
 
\textbf{Limitations.} The analysis reveals substantial inconsistency in existing MTD evaluations. Security-oriented metrics such as ASR and RA are widely reported, whereas operational metrics critical for deployment decisions are often omitted or only qualitatively discussed. In particular, controller overhead is discussed in nearly all studies but rarely quantified using concrete CPU or memory measurements. Similarly, throughput degradation and data exposure remain largely underexplored. These inconsistencies make fair comparison across MTD studies difficult and motivate the need for a unified composite evaluation framework that jointly captures security effectiveness and performance trade-offs under three attack scenarios as mentioned in Section \ref{sec:attack-scenarios}.

\subsubsection{Composite Scoring Formulation}
\label{sec:composite_formulation}
\noindent
While the metrics above provide useful low-level insights, individually they do not capture MTD deployment effectiveness, mutation-interval trade-offs, or defender--attacker operational balance. To address this limitation, we formulate a composite evaluation framework that jointly analyzes attacker disruption and operational overhead across different PR-MTD configurations and attack scenarios.

Our framework consists of three analytical components aligned with these evaluation objectives: (i) a deployment effectiveness formulation based on aggregate security and operational metrics, (ii) mutation-interval trade-off analysis based on attack disruption metrics, and (iii) defender--attacker utility analysis across different PR-MTD configurations.
Table~\ref{tab:composite_framework_overview} summarizes the three metrics and their interpretation.

\noindent\textbf{\textit{Metric 1: Deployment Effectiveness Ratio ($E$).}}
\label{sec:effectiveness_ratio}
This metric quantifies PR-MTD deployment effectiveness by comparing overall attacker disruption against the operational overhead introduced by mutation. We first derive the \textit{Attack Disruption Index (ADI)}, which captures attacker-facing disruption caused by PR-MTD, and the \textit{Operational Cost Index (OCI)}, which captures defender-side operational overhead.

For each index, we compute its relative change under MTD by dividing the MTD measurement by its corresponding No-MTD baseline. 

\textit{Attack Disruption Index (ADI)}: Security-oriented ratios are aggregated into the ADI using the geometric mean:
\begin{equation}
\text{ADI} = \left(\prod_{j=1}^{n_s} \frac{s^{\text{MTD}}_j}{s^{\text{NoMTD}}_j}\right)^{1/n_s}
\label{eq:sdi}
\end{equation}
where $s_j$ denotes the $j$-th security metric and $n_s$ is the number of security metrics. ADI incorporates seven security ratios, including exfiltration slowdown, scan-time increase, and attacker RTT variance across intrusion paths. Higher ADI values indicate stronger attacker disruption.

\textit{Operational Cost Index (OCI)}: Operational overhead metrics are similarly aggregated:
\begin{equation}
\text{OCI} = \left(\prod_{k=1}^{n_c} \frac{c^{\text{MTD}}_k}{c^{\text{NoMTD}}_k}\right)^{1/n_c}
\label{eq:oci}
\end{equation}
where $c_k$ denotes the $k$-th operational metric and $n_c$ is the number of cost metrics. OCI incorporates controller overhead, service latency, and inverse throughput across communication paths. Lower OCI values indicate lower operational overhead. For throughput, the ratio is inverted so that throughput degradation contributes positively to cost.

We use the geometric mean because the formulation is ratio-based and less sensitive to disproportionate influence from individual metrics.

\noindent\textbf{Effectiveness Ratio ($E$).}
The overall effectiveness ratio is computed as:
\begin{equation}
E = \frac{\text{ADI}}{\text{OCI}}
\label{eq:effectiveness}
\end{equation}

\textbf{Interpretation}. Higher $E$ values indicate that attacker disruption outweighs operational overhead. Because $E$ relies on normalized metric ratios without manually assigned weights, it can be consistently applied across different PR-MTD configurations and experimental settings.

\noindent\textbf{\textit{Metric 2: Mutation-Interval Sensitivity Analysis.}}
\label{sec:interval_metrics}
\noindent While the deployment effectiveness ratio quantifies the overall security--cost trade-off of PR-MTD, mutation-interval sensitivity analysis provides an additional perspective on how mutation speed influences attacker disruption. To analyze mutation-interval trade-offs, we model three interval-dependent existing  security metrics: Reconnaissance Accuracy (RA), Attack Success Rate (ASR), and Attack Completion Time (ACT). All three metrics are governed by the same principle: if the attacker requires more time to scan and exploit the network than the configured mutation interval, previously collected reconnaissance information becomes invalid before it can be used.

\noindent\textit{Reconnaissance Accuracy (RA).}
RA represents the probability that attacker reconnaissance remains valid when the exploit is launched. Let $T_{\text{recon}}$ denote reconnaissance time and $T_{\text{exploit}}$ denote exploit preparation and execution time. If the mutation interval $T$ is shorter than their combined duration, the forwarding path changes before the attacker can act:
\begin{equation}
\text{RA}(T) = \max!\left(0,;\frac{T - T_{\text{recon}} - T_{\text{exploit}}}{T}\right)
\label{eq:ra}
\end{equation}
Under static routing (No-MTD), $\text{RA}=1$. Faster mutation intervals reduce RA by invalidating previously collected network information.

\noindent\textit{Attack Success Rate (ASR).}
ASR models the probability of successfully completing a multi-stage intrusion campaign. In our enterprise topology, the attacker must successfully traverse two internal transitions (Web$\rightarrow$App and App$\rightarrow$DB), each requiring valid reconnaissance. Assuming independent success across both stages:
\begin{equation}
\widehat{\text{ASR}}(T) = \text{RA}(T)^2
\label{eq:asr}
\end{equation}
This compounding effect causes even moderate reductions in RA to significantly reduce end-to-end attack success. For attack scenarios bypassing one internal transition (e.g., SSRF-based pivoting), ASR reduces directly to RA.

\noindent\textit{Attack Completion Time (ACT).}
ACT captures the expected duration of a complete intrusion campaign under repeated reconnaissance failures. Lower RA values require attackers to repeatedly rescan the network before obtaining valid path information:
\begin{equation}
\widehat{\text{ACT}}(T) = \frac{T_{\text{scan}}}{\text{RA}(T)} + T_{\text{exfil}}
\label{eq:act}
\end{equation}
where $T_{\text{scan}}$ denotes the aggregate scan duration across attack paths and $T_{\text{exfil}}$ represents measured exfiltration time.

\textbf{Interpretation.} When the mutation interval becomes shorter than the attacker’s reconnaissance-plus-exploitation window ($T < T_{\text{recon}} + T_{\text{exploit}}$), $\text{RA}(T)$ and $\widehat{\text{ASR}}(T)$ decrease while $\widehat{\text{ACT}}(T)$ increases.

\noindent\textbf{\textit{Metric 3: Defender-Attacker Balance ($\Delta U$)}.}
\label{sec:utility_scores } While the previous formulations quantify deployment effectiveness and mutation-interval trade-offs, defender--attacker utility analysis provides additional insight into the relative operational advantage under different PR-MTD configurations.Accordingly,we compute utility scores from both defender and attacker perspectives and compare them through a net utility formulation.

Before aggregation, all metrics are normalized using min--max scaling to ensure comparability across heterogeneous units (e.g., seconds, percentages, and throughput). For metrics where lower values are preferable (e.g., latency and CPU utilization), the normalized values are inverted so that higher values consistently indicate more favorable outcomes. 

\textit{Defender Utility ($U_{\text{def}}$)}. 
The defender utility score combines three security-oriented metrics and three performance-oriented metrics:
\begin{equation}
\begin{split}
  U_{\text{def}} ={} & w_{1}\,\widehat{\text{ACT}}_n + w_{2}\,(1 - \widehat{\text{ASR}}_n) + w_{3}\,\widehat{\text{EXFIL}}_n \\
  & + w_{4}\,(1 - \widehat{\text{LAT}}_n) + w_{5}\,(1 - \widehat{\text{CPU}}_n) + w_{6}\,\widehat{\text{TP}}_n
\end{split}
\label{eq:udef}
\end{equation}
where $\widehat{\cdot}n$ denotes normalized values scaled between 0 and 1. The weighting scheme prioritizes security effectiveness (0.65 total weight) while incorporating operational performance (0.35 total weight) to reflect practical deployment considerations. \textit{Higher $U_{\text{def}}$ values indicate stronger defensive advantage}

\textit{Attacker Utility ($U_{\text{att}}$).}
The attacker utility score captures the attacker’s remaining operational capability:
\begin{equation}
U_{\text{att}} = w_7,\widehat{\text{ASR}}_n + w_8,(1 - \widehat{\text{ACT}}_n) + w_9,\widehat{\text{RA}}n
\label{eq:uatt}
\end{equation}
where higher $U_{\text{att}}$ values indicate more successful, faster, and reconnaissance-consistent attacks.

\noindent \textbf{Defender-Attacker Balance ($\Delta U$)}.
The overall defender--attacker balance is computed as: $\Delta U = U_{\text{def}} - U_{\text{att}}$

\textbf{Interpretation.} $\Delta U > 0$ indicates that the defender holds the operational advantage under the evaluated PR-MTD configuration, whereas $\Delta U < 0$ favors the attacker.

\begin{table}[!t]
\centering
\caption{Composite evaluation metrics overview.}
\label{tab:composite_framework_overview}
\footnotesize
\setlength{\tabcolsep}{3pt}
\renewcommand{\arraystretch}{1.12}
\begin{tabularx}{\columnwidth}{@{}l X X@{}}
\toprule
\textbf{Metric} & \textbf{Purpose} & \textbf{Interpretation} \\
\midrule

\makecell[l]{\textbf{M1:} Deployment \\ Effectiveness ($E$)}
&
Combines attacker disruption (ADI) and operational cost (OCI):
$E = \text{ADI}/\text{OCI}$
&
$E > 1$: deployment beneficial. Higher $E$ = stronger effectiveness.
\\

\midrule

\makecell[l]{\textbf{M2:} Mutation \\ Sensitivity}
&
Analyzes how RA, ASR, and ACT vary across mutation intervals.
&
Shorter mutation intervals reduce RA/ASR and increase ACT.
\\

\midrule

\makecell[l]{\textbf{M3:} Defender-- \\ Attacker Balance ($\Delta U$)}
&
Computes defender and attacker utility balance:
$\Delta U = U_{\text{def}} - U_{\text{att}}$
&
$\Delta U > 0$: defender advantage. $\Delta U < 0$: attacker advantage.
\\

\bottomrule
\end{tabularx}
\end{table}

\section{Results}
\label{sec:eval}

\noindent
This section presents the measured security, performance, and composite evaluation results generated using the \sys{} framework across three attack scenarios (\textbf{S1}:RCE, \textbf{S2}:SSRF, and \textbf{S3}:Credential-Based Compromise) under one baseline configuration without defense (No-MTD) and three PR-MTD configurations (MTD10s, MTD20s, and MTD30s). We first present security evaluation results (Section~\ref{sec:security_results}), followed by performance analysis (Section~\ref{sec:performance_results}), and finally the composite evaluation results (Section~\ref{sec:composite}) followed by takeaways for each.


\subsection{Security Results}
\label{sec:security_results}

\noindent Table~\ref{tab:security_all_scenarios} reports all security metrics across the three attack scenarios (S1, S2, S3) and four conditions (No-MTD, MTD10\,s, MTD20\,s, MTD30\,s). Results are grouped by metric category below.

\textbf{Reconnaissance and Attack Success.} At MTD10\,s, \textit{RA} drops to 20\% in S1 and S2 and to 40\% in S3, with compound \textit{ASR} falling to 4\%, 20\%, and 16\% respectively. S1 shows the lowest ASR because both internal transitions require independent scanning, while S2 retains higher ASR because SSRF bypasses one scanning step. \textit{ACT} increases from 43--63\,s under No-MTD to 311\,s (S1), 210\,s (S2), and 160\,s (S3) at MTD10\,s due to repeated reconnaissance failures. At MTD30\,s, RA recovers to 73--80\% and ASR rises to 54--73\%, leaving viable attack paths in every scenario.

\textbf{Network Unpredictability.} This measures how much the attacker's repeated ping results vary on each path, higher variance means the attacker cannot reliably distinguish network tiers or confirm host reachability. The largest variance increase is on Web$\to$App in S1 and S3 (up to 32$\times$ at MTD10\,s) where the attacker actively probes the path. S2 shows lower Web$\to$App variance (12.8$\times$) because SSRF uses the web server's existing connection instead of direct probing. Att$\to$Web variance reaches 15.6$\times$ in S1 and S2 but only 9.4$\times$ in S3 because credential-based access reduces probing on that path. App$\to$DB variance is consistent across all scenarios (2.5$\times$ at MTD10\,s).

\textbf{Scan Time.} Under No-MTD, scan durations vary by path (15.2--16.8\,s), enabling timing-based tier identification. Under PR-MTD, S1 scan times converge toward $\approx$20\,s across all paths, removing this timing signature. In S2, Att$\to$Web and App$\to$DB converge similarly, but 
Web$\to$App drops to $\approx$6\,s because SSRF reduces active scanning on that path. In S3, Att$\to$Web drops to $\approx$10\,s because credential access eliminates full port scanning, while Web$\to$App and App$\to$DB still converge toward $\approx$20\,s.

\textbf{Traceroute Total RTT.} While the previous metric measures variation across repeated probes, this measures the total time a packet takes to travel from source to destination across all hops on a path. At MTD10\,s, total RTT increases on some paths (Att$\to$Web rises from 0.130\,ms to 0.153\,ms) because frequent flow-rule changes add forwarding delay. At MTD30\,s, total RTT drops below the No-MTD baseline as longer-lived rules benefit from OVS caching: Web$\to$App falls from 0.144\,ms to 0.085\,ms and App$\to$DB from 0.139\,ms to 0.087\,ms in S1. S2 shows slightly higher Web$\to$App RTT at MTD30\,s (0.098\,ms) due to server-side overhead from SSRF. S3 shows lower Att$\to$Web RTT (0.080\,ms) from faster authenticated access but slightly higher App$\to$DB (0.091\,ms).

\textbf{Data Exposure.} Exfiltration time increases from 2.456\,s under No-MTD to 12.66--14.83\,s at MTD10\,s across all scenarios. S3 has the longest delay (14.83\,s) because authenticated sessions require re-establishment after each path mutation, while S2 has the shortest (12.66\,s) because server-side connections partially persist across path changes. At MTD30\,s, exfiltration still takes 7.0--8.2\,s, confirming that even slower mutation intervals disrupt active data transfer.

\begin{tcolorbox}[
colback=gray!20,
colframe=black!40,
boxrule=1pt,
arc=1mm,
left=1mm,
right=1mm,
top=0.5mm,
bottom=0.5mm
]
\textbf{Security Evaluation Takeaway.}
MTD10\,s provides the strongest attacker disruption, reducing ASR to 4--20\%, extending ACT to 160--311\,s, and significantly increasing network unpredictability and exfiltration delay. S1 (RCE) is most affected because both internal transitions require reconnaissance, whereas S2 (SSRF) is least affected due to bypassing one scanning stage. At MTD30\,s, ASR recovers to 54--73\%, indicating that shorter mutation intervals are critical for effective PR-MTD defense.

\end{tcolorbox}


\begin{table*}[t]
\centering
\caption{Security metric results across three evaluated attack scenarios under No-MTD and PR-MTD (10s, 20s and 30s) configurations. S1 (RCE) requires reconnaissance across both internal transitions, S2 (SSRF) bypasses one reconnaissance stage through server-side forwarding, and S3 (Credential-Based) reduces initial reconnaissance through authenticated access.}
\label{tab:security_all_scenarios}
\footnotesize
\setlength{\tabcolsep}{3.5pt}
\renewcommand{\arraystretch}{1.1}
\begin{tabular}{l c | r r r r | r r r r | r r r r}
\hline
& & \multicolumn{4}{c|}{\textbf{S1: RCE-driven}} & \multicolumn{4}{c|}{\textbf{S2: SSRF-driven}} & \multicolumn{4}{c}{\textbf{S3: Credential-driven}} \\
\textbf{Metric} & \textbf{Path} & \textbf{No-MTD} & \textbf{MTD10\,s} & \textbf{MTD20\,s} & \textbf{MTD30\,s} & \textbf{No-MTD} & \textbf{MTD10\,s} & \textbf{MTD20\,s} & \textbf{MTD30\,s} & \textbf{No-MTD} & \textbf{MTD10\,s} & \textbf{MTD20\,s} & \textbf{MTD30\,s} \\
\hline
\multicolumn{14}{l}{\textit{Reconnaissance \& Attack Success}} \\
RA (\%) & -- & 100 & \textbf{20.0} & 60.0 & 73.3 & 100 & \textbf{20.0} & 60.0 & 73.3 & 100 & \textbf{40.0} & 70.0 & 80.0 \\
ASR (\%) & -- & 100 & \textbf{4.0} & 36.0 & 53.8 & 100 & \textbf{20.0} & 60.0 & 73.3 & 100 & \textbf{16.0} & 49.0 & 64.0 \\
ACT (s) & -- & 63.2 & \textbf{311.1} & 108.7 & 90.3 & 42.9 & \textbf{209.5} & 74.9 & 62.6 & 63.2 & \textbf{160.0} & 94.9 & 84.1 \\
\hline
\multicolumn{14}{l}{\textit{Network Unpredictability (RTT variance $\times$)}} \\
\multirow{3}{*}{\makecell[l]{RTT\\Variance}} & A$\to$W & 1.0 & 15.6 & 12.5 & 10.4 & 1.0 & 15.6 & 12.5 & 10.4 & 1.0 & 9.4 & 7.5 & 6.2 \\
& W$\to$A & 1.0 & 32.0 & 25.6 & 21.3 & 1.0 & 12.8 & 10.2 & 8.5 & 1.0 & 32.0 & 25.6 & 21.3 \\
& A$\to$D & 1.0 & 2.5 & 2.0 & 1.7 & 1.0 & 2.5 & 2.0 & 1.7 & 1.0 & 2.5 & 2.0 & 1.7 \\
\hline
\multicolumn{14}{l}{\textit{Scan Time (s)}} \\
\multirow{3}{*}{\makecell[l]{Scan\\Time}} & A$\to$W & 15.16 & 20.57 & 20.37 & 20.17 & 15.16 & 20.57 & 20.37 & 20.17 & 15.16 & 10.29 & 10.19 & 10.09 \\
& W$\to$A & 16.01 & 20.73 & 20.52 & 20.32 & 16.01 & 6.22 & 6.16 & 6.10 & 16.01 & 20.73 & 20.52 & 20.32 \\
& A$\to$D & 16.82 & 20.63 & 20.43 & 20.23 & 16.82 & 20.63 & 20.43 & 20.23 & 16.82 & 20.63 & 20.43 & 20.23 \\
\hline
\multicolumn{14}{l}{\textit{Traceroute Total RTT (ms)}} \\
\multirow{3}{*}{\makecell[l]{Trace\\RTT}} & A$\to$W & 0.130 & 0.153 & 0.115 & 0.094 & 0.130 & 0.153 & 0.115 & 0.094 & 0.130 & 0.130 & 0.098 & 0.080 \\
& W$\to$A & 0.144 & 0.120 & 0.099 & 0.085 & 0.144 & 0.138 & 0.114 & 0.098 & 0.144 & 0.120 & 0.099 & 0.085 \\
& A$\to$D & 0.139 & 0.126 & 0.095 & 0.087 & 0.139 & 0.126 & 0.095 & 0.087 & 0.139 & 0.132 & 0.100 & 0.091 \\
\hline
\multicolumn{14}{l}{\textit{Data Exposure}} \\
Exfiltration (s) & -- & 2.456 & \textbf{13.65} & 10.09 & 7.55 & 2.456 & \textbf{12.66} & 9.36 & 7.00 & 2.456 & \textbf{14.83} & 10.97 & 8.20 \\
\hline
\end{tabular}
\par\vspace{0.3em}
\scriptsize{
\textbf{Path} = Network Path,
\textbf{A$\to$W} = Attacker$\to$Web,
\textbf{W$\to$A} = Web$\to$App,
\textbf{A$\to$D} = App$\to$DB.
}
\end{table*}
\vspace{0.3cm}
\subsection{Performance Results}
\label{sec:performance_results}
\noindent Unlike the security metrics, which vary across attack scenarios due to differences in reconnaissance and attack progression, the performance metrics are determined entirely by network-layer behavior. The ONOS controller applies identical path-randomization schedules, forwarding rules, and mutation logic across all scenarios. Consequently, the performance results in Table~\ref{tab:performance_summary} apply uniformly to S1:RCE, S2:SSRF, and S3:Credential-Based Compromise.

\textbf{Service Latency.}
Service latency varies across network segments and mutation intervals. Compared with No-MTD, Att$\to$Web latency increases from 5.20ms to 9.13ms, 7.89ms, and 7.32ms at MTD10s, MTD20s, and MTD30s, respectively. A similar trend is observed for Web$\to$App, where latency rises from 1.92ms to 4.72ms, 4.08ms, and 3.78ms. In contrast, App$\to$DB latency decreases from 0.69ms to 0.30ms, 0.26ms, and 0.24ms across the same mutation intervals. 
Shorter intervals produce slightly higher latency on external paths due to more frequent flow-rule transitions, while longer intervals benefit from OVS caching. Prior work has reported stable RTT under path mutation~\cite{Abdelkhalek2022}, but latency improvement through hop reduction has not been documented.

\textbf{Throughput.} PR-MTD improves throughput on every path. At MTD30\,s, Att$\to$Web increases from 13.7 to 14.6\,Gbits/s (+6.6\%), Web$\to$App from 13.6 to 14.0\,Gbits/s (+2.9\%), and App$\to$DB from 13.9 to 18.2\,Gbits/s (+30.9\%). At MTD10\,s, improvements are smaller,  14.2, 13.8, and 16.5\,Gbits/s respectively, because more frequent mutations reduce the caching benefit. At MTD20\,s, values fall between: 14.4, 13.9, and 17.3\,Gbits/s. Among all representative studies surveyed, only three report throughput and none report improvement under MTD.\newline
\indent \textbf{Controller Overhead.} ONOS shows the primary overhead increase under PR-MTD. At MTD10\,s, CPU reaches 14.90\% (+7.34\,pp from baseline 7.56\%) and memory reaches 36.69\% (+15.71\,pp from baseline 20.98\%). At MTD20\,s, CPU is 12.70\% (+5.14\,pp) and memory is 32.76\% (+11.78\,pp). At MTD30\,s, CPU is 11.23\% (+3.67\,pp) and memory is 30.80\% (+9.82\,pp). Overhead decreases at longer intervals because fewer flow-rule updates are needed. Despite the percentage increase at MTD10\,s, absolute values remain moderate. To our knowledge, this is the first quantitative controller-overhead measurement for path-randomisation MTD, as all prior studies report overhead only qualitatively.\newline
\indent \textbf{Network Structure.} Under No-MTD, internal paths (Att$\to$Web, Web$\to$App and App$\to$DB) traverse 2 hops. Under all PR-MTD intervals, the ONOS controller installs direct forwarding rules that bypass the intermediate switch, reducing all internal paths to 2 hops. This structural change is consistent across MTD10\,s, MTD20\,s, and MTD30\,s, and explains both the App$\to$DB latency reduction and throughput improvement described above.\newline
\indent \textbf{Service Availability.} All 10 experimental runs under both No-MTD and PR-MTD complete with zero service interruptions across all three intervals. No connectivity loss or recovery delay was observed during any path mutation event. This contrasts with VM-based MTD that reports 11$\pm$3\,s service lapses per reincarnation cycle~\cite{Ahmed2016}.

\begin{table}[t]
\centering
\caption{Performance metric results comparing the No-MTD baseline with path-randomization configurations (MTD10/20/30s).}
\label{tab:performance_summary}
\small
\setlength{\tabcolsep}{3pt}
\renewcommand{\arraystretch}{1.1}
\begin{tabularx}{\columnwidth}{@{}X r r r r@{}}
\hline
\textbf{Metric} & \textbf{No-MTD} & \textbf{MTD10\,s} & \textbf{MTD20\,s} & \textbf{MTD30\,s} \\
\hline
\multicolumn{5}{@{}l}{\textit{\textbf{Service Latency (ms)}}} \\
Att$\to$Web & 5.20 & 9.13 & 7.89 & \textbf{7.32} \\
Web$\to$App & 1.92 & 4.72 & 4.08 & \textbf{3.78} \\
App$\to$DB & 0.69 & 0.30 & 0.26 & \textbf{0.24} \\
\hline
\multicolumn{5}{@{}l}{\textit{\textbf{Throughput (Gbits/s)}}} \\
Att$\to$Web & 13.7 & 14.2 & 14.4 & \textbf{14.6} \\
Web$\to$App & 13.6 & 13.8 & 13.9 & \textbf{14.0} \\
App$\to$DB & 13.9 & 16.5 & 17.3 & \textbf{18.2} \\
\hline
\multicolumn{5}{@{}l}{\textit{\textbf{Controller Overhead}}} \\
ONOS CPU (\%) & 7.56 & 14.90 & 12.70 & \textbf{11.23} \\
ONOS Memory (\%) & 20.98 & 36.69 & 32.76 & \textbf{30.80} \\
\hline
\multicolumn{5}{@{}l}{\textit{\textbf{Network Structure}}} \\
Hops (Att$\to$Web) & 2 & 2 & 2 & \textbf{2} \\
Hops (Web$\to$App) & 2 & 2 & 2 & \textbf{2} \\
Hops (App$\to$DB) & 2 & 2 & 2 & \textbf{2} \\
\hline
\multicolumn{5}{@{}l}{\textit{\textbf{Service Availability}}} \\
Interruptions & 0 & 0 & 0 & \textbf{0} \\
\hline

\end{tabularx}
\end{table}

\begin{tcolorbox}[
colback=gray!20,
colframe=black!40,
boxrule=1pt,
arc=1mm,
left=1mm,
right=1mm,
top=0.5mm,
bottom=0.5mm
]
\textbf{Performance Evaluation Takeaway.}
Contrary to the common assumption that stronger MTD mechanisms degrade network performance, PR-MTD improves throughput on all evaluated paths (up to +30.9\%) while reducing internal-path latency by up to 65\%. Although controller overhead increases under shorter mutation intervals, utilization remains moderate across all configurations (below 15\% CPU and 37\% memory) with zero service interruption. Among all evaluated configurations, MTD30\,s achieves the best operational performance due to longer-lived flow rules benefiting from OVS fast-path caching, resulting in the highest throughput and lowest latency measurements.

\end{tcolorbox}

\begin{table}[t]
\centering
\caption{Composite evaluation results across all PR-MTD configurations and attack scenarios. \textbf{Metric 1}: ADI and $E$ ($ = \text{ADI}/\text{OCI}$, OCI $= 1.049$). \textbf{Metric 2}: RA, ASR, and ACT per interval. \textbf{Metric 3}: $U_{\text{def}}$, $U_{\text{att}}$, and $\Delta U$ (globally normalised). Bold = strongest defender outcome. Grey-shaded rows = defender-favourable interval.}
\label{tab:composite_all}
\footnotesize
\setlength{\tabcolsep}{2pt}
\renewcommand{\arraystretch}{1.2}
\begin{tabular}{l | r r | r r | r r}
\hline
& \multicolumn{2}{c|}{\textbf{S1: RCE-driven}} & \multicolumn{2}{c|}{\textbf{S2: SSRF-driven}} & \multicolumn{2}{c}{\textbf{S3: Credential-driven}} \\
\hline
\multicolumn{7}{l}{\cellcolor{gray!15}\textbf{Metric 1: Deployment Effectiveness Ratio}} \\
\hline
\textbf{Condition} & \textbf{ADI} & \textbf{$E$} & \textbf{ADI} & \textbf{$E$} & \textbf{ADI} & \textbf{$E$} \\
\hline
\rowcolor{gray!8}
MTD10\,s & \textbf{3.95} & \textbf{3.76} & \textbf{2.89} & \textbf{2.75} & \textbf{3.37} & \textbf{3.21} \\
MTD20\,s & 3.42 & 3.26 & 2.50 & 2.38 & 2.92 & 2.78 \\
MTD30\,s & 3.03 & 2.89 & 2.21 & 2.11 & 2.58 & 2.46 \\
OCI & \multicolumn{6}{c}{1.049 (constant across all conditions)} \\
\hline
\multicolumn{7}{l}{\cellcolor{gray!15}\textbf{Metric 2: Mutation-Interval Sensitivity}} \\
\hline
\textbf{Condition} & \textbf{RA (\%)} & \textbf{ASR (\%)} & \textbf{RA (\%)} & \textbf{ASR (\%)} & \textbf{RA (\%)} & \textbf{ASR (\%)} \\
\hline
No-MTD & 100 & 100 & 100 & 100 & 100 & 100 \\
\rowcolor{gray!8}
MTD10\,s & \textbf{20.0} & \textbf{4.0} & \textbf{20.0} & \textbf{20.0} & \textbf{40.0} & \textbf{16.0} \\
MTD20\,s & 60.0 & 36.0 & 60.0 & 60.0 & 70.0 & 49.0 \\
MTD30\,s & 73.3 & 53.8 & 73.3 & 73.3 & 80.0 & 64.0 \\
\hline
\textbf{Condition} & \textbf{ACT (s)} & & \textbf{ACT (s)} & & \textbf{ACT (s)} & \\
\hline
No-MTD & 63.2 & & 42.9 & & 63.2 & \\
\rowcolor{gray!8}
MTD10\,s & \textbf{311.1} & & \textbf{209.5} & & \textbf{160.0} & \\
MTD20\,s & 108.7 & & 74.9 & & 94.9 & \\
MTD30\,s & 90.3 & & 62.6 & & 84.1 & \\
\hline
\multicolumn{7}{l}{\cellcolor{gray!15}\textbf{Metric 3: Defender-Attacker Balance}} \\
\hline
\textbf{Condition} & $U_{\text{def}}$ / $U_{\text{att}}$ & $\Delta U$ & $U_{\text{def}}$ / $U_{\text{att}}$ & $\Delta U$ & $U_{\text{def}}$ / $U_{\text{att}}$ & $\Delta U$ \\
\hline
No-MTD & 0.26 / 0.98 & $-$0.72 & 0.25 / 1.00 & $-$0.75 & 0.25 / 1.00 & $-$0.74 \\
\rowcolor{gray!8}
MTD10\,s & \textbf{0.69 / 0.00} & \textbf{+0.69} & \textbf{0.57 / 0.15} & \textbf{+0.42} & \textbf{0.51 / 0.33} & \textbf{+0.18} \\
MTD20\,s & 0.42 / 0.51 & $-$0.09 & 0.33 / 0.64 & $-$0.31 & 0.37 / 0.64 & $-$0.26 \\
MTD30\,s & 0.37 / 0.66 & $-$0.29 & 0.29 / 0.76 & $-$0.47 & 0.33 / 0.75 & $-$0.41 \\
\hline
\end{tabular}
\par\vspace{0.3em}
\raggedright\scriptsize{OCI = 1.049 across all conditions (4.9\% aggregate cost). $E = \text{ADI}/\text{OCI}$; all $E > 1$ confirms net-beneficial deployment. $\Delta U$ normalised globally across all 12 condition--scenario combinations. Grey-shaded rows indicate MTD10\,s, the only interval where the defender wins in every scenario.}
\end{table}

\subsection{Composite Evaluation Results}
\label{sec:composite}

\noindent
The security and performance results above evaluate individual metrics independently. Here, we apply the composite formulations from Section~\ref{sec:composite_formulation} to derive unified conclusions on deployment effectiveness, mutation-interval trade-offs, and defender--attacker operational balance. Table~\ref{tab:composite_all} summarizes the composite evaluation results across all attack scenarios and PR-MTD configurations.

\textbf{{Deployment Effectiveness.}}
Table~\ref{tab:composite_all}, Metric 1 reports ADI and $E$ at each interval for all three scenarios. Every $E$ value is well above 1, confirming that path randomisation is net-beneficial at every interval and for every attack type. The lowest combination (S2 at MTD30\,s, $E = 2.11$) still delivers twice the security benefit per unit of cost. MTD10\,s produces the highest $E$ in every scenario, 3.76 for S1, 2.75 for S2, and 3.21 for S3, because more frequent path changes increase RTT variance and exfiltration delay. The scenario ranking is consistent across all intervals: S1 (RCE) always has the highest $E$ because the attacker scans all paths and navigates both internal transitions, S2 (SSRF) the lowest because the web server's trusted connection reduces scanning exposure, and S3 (Credential) falls between. OCI remains constant at 1.049 (4.9\% aggregate overhead) across all scenarios and mutation intervals because the network layer does not distinguish between attack techniques (see Appendix Table~\ref{tab:oci_breakdown}). 

Table \ref{tab:adi_ratios} provides a detailed breakdown of the individual ratios underlying the ADI and OCI values reported in Table ~\ref{tab:composite_all}. For each interval and scenario, the table presents the seven security ratios used to compute ADI, with each ratio calculated as the MTD value divided by the corresponding No-MTD baseline. The ADI is then obtained as the geometric mean of these seven ratios, providing an aggregate measure of the security benefit achieved through path randomization.

\begin{table*}[t]
\centering
\caption{ADI ratio breakdown across all mutation intervals and three scenarios. Each ratio is computed as MTD value / No-MTD value; higher values indicate greater attacker disruption. ADI is the geometric mean of the seven ratios. $E = \text{ADI}/\text{OCI}$, where $\text{OCI}=1.049$.}

\label{tab:adi_ratios}
\footnotesize
\setlength{\tabcolsep}{3.5pt}
\renewcommand{\arraystretch}{1.15}
\begin{tabular}{l | r r r | r r r | r r r}
\hline
& \multicolumn{3}{c|}{\textbf{S1: RCE}} & \multicolumn{3}{c|}{\textbf{S2: SSRF}} & \multicolumn{3}{c}{\textbf{S3: Credential}} \\
\textbf{ADI Ratio} & \textbf{MTD10\,s} & \textbf{MTD20\,s} & \textbf{MTD30\,s} & \textbf{MTD10\,s} & \textbf{MTD20\,s} & \textbf{MTD30\,s} & \textbf{MTD10\,s} & \textbf{MTD20\,s} & \textbf{MTD30\,s} \\
\hline
Exfiltration & 5.56 & 4.11 & 3.07 & 5.15 & 3.81 & 2.85 & 6.04 & 4.47 & 3.34 \\
Scan --- Att$\to$Web & 1.36 & 1.34 & 1.33 & 1.36 & 1.34 & 1.33 & 0.68 & 0.67 & 0.67 \\
Scan --- Web$\to$App & 1.29 & 1.28 & 1.27 & 0.39 & 0.38 & 0.38 & 1.29 & 1.28 & 1.27 \\
Scan --- App$\to$DB & 1.23 & 1.21 & 1.20 & 1.23 & 1.21 & 1.20 & 1.23 & 1.21 & 1.20 \\
RTT var --- Att$\to$Web & 15.6 & 12.5 & 10.4 & 15.6 & 12.5 & 10.4 & 9.4 & 7.5 & 6.2 \\
RTT var --- Web$\to$App & 32.0 & 25.6 & 21.3 & 12.8 & 10.2 & 8.5 & 32.0 & 25.6 & 21.3 \\
RTT var --- App$\to$DB & 2.5 & 2.0 & 1.7 & 2.5 & 2.0 & 1.7 & 2.5 & 2.0 & 1.7 \\
\hline
\textbf{ADI} & \textbf{3.95} & 3.42 & 3.03 & \textbf{2.89} & 2.50 & 2.21 & \textbf{3.37} & 2.92 & 2.58 \\
\textbf{$E$} & \textbf{3.76} & 3.26 & 2.89 & \textbf{2.75} & 2.38 & 2.11 & \textbf{3.21} & 2.78 & 2.46 \\
\hline
\end{tabular}
\end{table*}

\textbf{Mutation-Interval Trade-offs.} Table~\ref{tab:composite_all}, Metric 2 reports RA, ASR, and ACT at each interval. The sharp drop between MTD20\,s and MTD10\,s is visible in every scenario. In S1, ASR falls from 36\% to 4\%, the attacker goes from succeeding one-third of the time to almost never. In S2, the drop is from 60\% to 20\%, less dramatic because SSRF skips one scanning step but still a major reduction. In S3, the drop is from 49\% to 16\%, with faster credential access partially cushioning the effect. ACT follows the same pattern: At MTD10\,s, attacks take 311\,s (S1), 210\,s (S2), and 160\,s (S3), compared to 43--63\,s under No-MTD. At MTD30\,s, ASR recovers to 54--73\% and ACT drops to 63--90\,s across all scenarios, confirming that slower mutation leaves viable attack paths regardless of technique. 

Additional sensitivity analysis across varying attacker probe and exploitation speeds is provided in Table~\ref{tab:asr_sensitivity}. The table reports compound ASR under different combinations of attacker probe and exploitation speeds, showing how changes in attacker capabilities affect the observed attack success rate. S1 and S2 use $T_{\text{exploit}}$ as defined above, while S3 uses $T_{\text{exploit}} - 2$\,s to account for faster credential login. The highlighted row corresponds to the attacker parameters used in our evaluation.

\begin{table*}[t]
\centering
\caption{Compound ASR (\%) under varying attacker probe and exploitation speeds across all scenarios. MTD10\,s consistently yields the lowest ASR. The highlighted row indicates the attacker parameters used in the evaluation.}

\label{tab:asr_sensitivity}
\footnotesize
\setlength{\tabcolsep}{4pt}
\renewcommand{\arraystretch}{1.15}
\begin{tabular}{c c | r r r | r r r | r r r}
\hline
& & \multicolumn{3}{c|}{\textbf{S1: RCE}} & \multicolumn{3}{c|}{\textbf{S2: SSRF}} & \multicolumn{3}{c}{\textbf{S3: Credential}} \\
$T_{\text{recon}}$ & $T_{\text{exploit}}$ & \textbf{MTD10\,s} & \textbf{MTD20\,s} & \textbf{MTD30\,s} & \textbf{MTD10\,s} & \textbf{MTD20\,s} & \textbf{MTD30\,s} & \textbf{MTD10\,s} & \textbf{MTD20\,s} & \textbf{MTD30\,s} \\
\hline
2\,s & 3\,s & 25.0 & 56.2 & 69.4 & 50.0 & 75.0 & 83.3 & 49.0 & 72.2 & 81.0 \\
2\,s & 5\,s & 9.0 & 42.3 & 58.8 & 30.0 & 65.0 & 76.7 & 25.0 & 56.2 & 69.4 \\
\rowcolor{gray!15}
3\,s & 5\,s & 4.0 & 36.0 & 53.8 & 20.0 & 60.0 & 73.3 & 16.0 & 49.0 & 64.0 \\
3\,s & 8\,s & 0.0 & 20.2 & 40.1 & 0.0 & 45.0 & 63.3 & 1.0 & 30.3 & 49.0 \\
5\,s & 5\,s & 0.0 & 25.0 & 44.4 & 0.0 & 50.0 & 66.7 & 4.0 & 36.0 & 53.8 \\
\hline
\end{tabular}
\end{table*}

\textbf{Defender-Attacker Operational Balance.}
Table~\ref{tab:composite_all}, Metric 3 reports $U_{\text{def}}$, $U_{\text{att}}$, and $\Delta U$ at each interval. Scores are normalised globally across all scenarios and conditions, so $\Delta U$ values are directly comparable.
 
Without MTD, the attacker dominates in every scenario ($\Delta U = -0.72$ to $-0.75$). At MTD10\,s, the defender wins in all three scenarios but with different margins. S1 (RCE) gives the strongest advantage ($\Delta U = +0.69$): attacker utility collapses to 0.00 because ASR drops to 4\% and ACT exceeds 5 minutes. S2 (SSRF) gives a moderate advantage ($\Delta U = +0.42$): the attacker retains some capability ($U_{\text{att}} = 0.15$) because the web server's trusted connection bypasses one scanning step. S3 (Credential) gives the slimmest advantage ($\Delta U = +0.18$): faster login raises attacker utility to 0.33. At MTD20\,s, the attacker regains the edge in every scenario ($\Delta U = -0.09$ to $-0.31$), and at MTD30\,s holds a clear advantage ($\Delta U = -0.29$ to $-0.47$), with S2 being the worst case for the defender. 

To assess the robustness of this result, Table~\ref{tab:weight_sensitivity_all} reports $\Delta U$ under three security--performance weight configurations. MTD10\,s yields a positive $\Delta U$ across all weight configurations and scenarios, confirming that the defender advantage persists despite changes in the relative weighting of security and performance.

\begin{table}[t]
\centering
\caption{$\Delta U$ under different security--performance weight splits across all scenarios. Bold values indicate defender advantage. MTD10\,s yields positive $\Delta U$ across all weight configurations and scenarios.}
\label{tab:weight_sensitivity_all}
\footnotesize
\setlength{\tabcolsep}{3pt}
\renewcommand{\arraystretch}{1.12}

\begin{tabular*}{\columnwidth}
{@{\extracolsep{\fill}} l l r r r r @{}}
\hline
\textbf{Weights} & \textbf{Scen.} & \textbf{No-MTD} &
\textbf{MTD10\,s} & \textbf{MTD20\,s} & \textbf{MTD30\,s} \\
\hline

\multirow{3}{*}{\makecell[l]{Sec 50\\Perf 50}}
& S1 & $-$0.62 & \textbf{+0.58} & $-$0.11 & $-$0.27 \\
& S2 & $-$0.65 & \textbf{+0.33} & $-$0.31 & $-$0.43 \\
& S3 & $-$0.64 & \textbf{+0.10} & $-$0.28 & $-$0.39 \\
\hline

\multirow{3}{*}{\makecell[l]{Sec 65\\Perf 35}}
& S1 & $-$0.72 & \textbf{+0.69} & $-$0.09 & $-$0.29 \\
& S2 & $-$0.75 & \textbf{+0.42} & $-$0.31 & $-$0.47 \\
& S3 & $-$0.74 & \textbf{+0.18} & $-$0.26 & $-$0.41 \\
\hline

\multirow{3}{*}{\makecell[l]{Sec 80\\Perf 20}}
& S1 & $-$0.82 & \textbf{+0.81} & $-$0.07 & $-$0.31 \\
& S2 & $-$0.85 & \textbf{+0.51} & $-$0.31 & $-$0.51 \\
& S3 & $-$0.84 & \textbf{+0.26} & $-$0.25 & $-$0.44 \\
\hline
\end{tabular*}
\end{table}

\begin{tcolorbox}[
colback=gray!20,
colframe=black!40,
boxrule=1pt,
arc=1mm,
left=1mm,
right=1mm,
top=0.5mm,
bottom=0.5mm
]
\textbf{Composite Evaluation Takeaway.}
PR-MTD remains deployment-beneficial across all evaluated scenarios and intervals, with effectiveness ratios ($E$) ranging from 2.11 to 3.76 under only 4.9\% operational overhead. MTD10\,s is the only configuration that consistently achieves positive defender advantage across all attack scenarios, with the strongest impact against reconnaissance-driven RCE attacks and the weakest against SSRF-based attacks that bypass one scanning stage. At MTD20\,s and MTD30\,s, attacker advantage re-emerges across all scenarios, highlighting that aggressive mutation intervals are critical for effective PR-MTD defense.
\end{tcolorbox}

\section{Discussion}
\label{sec:discussion}
The following findings translate the measured security, performance, and composite results into deployment-oriented insights for PR-MTD.

\noindent\textbf{Finding 1: Deployment-Ready Mutation Interval.}
Our results show that PR-MTD effectiveness is primarily governed by the relationship between mutation interval and attacker reconnaissance time. Across all scenarios, MTD10\,s achieves the highest deployment effectiveness ($E$), lowest attack success rates (4--20\%), and positive defender advantage ($\Delta U>0$). In contrast, MTD20\,s and MTD30\,s allow attackers sufficient time to reuse reconnaissance results, causing attack success to recover and defender advantage to disappear. Thus, effective PR-MTD deployment requires mutation intervals shorter than the attacker's reconnaissance-plus-exploitation window.

Table~\ref{tab:deployment_summary} summarizes the deployment-level outcome across all evaluated scenarios. MTD10\,s is consistently the highest-performing configuration, achieving the highest $E$ and positive defender advantage in every scenario.

\begin{table}[h]
\centering
\caption{Deployment outcome summary across all evaluated attack scenarios.}
\label{tab:deployment_summary}
\footnotesize
\setlength{\tabcolsep}{4pt}
\renewcommand{\arraystretch}{1.12}
\begin{tabularx}{\columnwidth}{@{}l >{\centering\arraybackslash}X >{\centering\arraybackslash}X >{\centering\arraybackslash}X@{}}
\hline
\textbf{Scenario} & \textbf{Best Interval} & \textbf{Highest $E$} & \textbf{$\Delta U > 0$} \\
\hline
S1 (RCE) & MTD10\,s & 3.76 & Yes \\
S2 (SSRF) & MTD10\,s & 2.75 & Yes \\
S3 (Credential) & MTD10\,s & 3.21 & Yes \\
\hline
\end{tabularx}
\end{table}

\noindent\textbf{Finding 2: Attack Structure Matters.}
Attack structure strongly influences PR-MTD effectiveness. RCE experiences the strongest disruption because each internal transition must be independently rediscovered after mutation. In contrast, SSRF partially bypasses this protection by leveraging trusted server-side communication, reducing the compounding disruption effect. This indicates that path randomization alone is insufficient against application-layer forwarding attacks and should be complemented by SSRF filtering and application-layer validation.

\noindent\textbf{Finding 3: Security Need Not Trade Off Against Performance.}
PR-MTD does not impose the performance penalties commonly assumed for MTD. Instead, throughput improves by up to 30.9\% and internal-path latency decreases through SDN flow reconfiguration. Although shorter intervals increase controller overhead, utilization remains moderate and no service interruption occurs across the evaluated configurations. These results indicate that SDN-enabled PR-MTD can strengthen security while maintaining practical network performance.

\indent\textbf{Broader Implication.}
\sys{} demonstrates the value of deployment-oriented MTD evaluation. Rather than reporting isolated security or performance metrics, it jointly analyzes attacker disruption, operational overhead, mutation sensitivity, and defender--attacker balance, enabling practitioners to identify not only whether PR-MTD is effective, but also which configuration is deployment-worthy under a given attack scenario.

\section{Limitations \& Future Work}
\label{sec:limit}
\noindent
Although \sys{} provides a controlled and reproducible framework for evaluating SDN-based PR-MTD techniques, several limitations remain. \textit{First}, the current implementation focuses on periodic path randomization and does not yet incorporate adaptive or event-driven mutation strategies, IP shuffling, deception-based defenses, or AI/ML and game-theoretic (GT) decision mechanisms. \textit{Second}, the evaluation is conducted on a simplified small enterprise-style topology designed for controlled experimentation and repeatable analysis rather than production-scale deployment. Larger and more heterogeneous environments may introduce additional challenges related to controller scalability, flow-rule coordination, and network stability. \textit{Third}, attacker behavior is modeled using predefined multi-stage attack workflows under limited attacker knowledge assumptions. Real-world adversaries may exhibit more adaptive, stealth-aware, or coordinated behaviors that are not fully represented in the current evaluation setting.

\indent Future work will extend \sys{} in several directions. Additional MTD mechanisms, including IP mutation, deception-based defenses, and AI/ML or GT-driven strategies, will be integrated to support broader comparative benchmarking. We also plan to explore adaptive mutation scheduling using reinforcement learning and attacker-aware decision policies to evaluate PR-MTD under dynamically evolving adversarial conditions~\cite{Moghaddam2024}. Finally, the framework will be evaluated on larger multi-controller enterprise topologies with diverse services and traffic patterns to further study scalability, coordination overhead, and generalizability of the proposed evaluation methodology.

\section{Conclusion}
\label{sec:conclusion}
\noindent In this paper, we presented \sys{}, an attacker-aware evaluation framework for benchmarking SDN-enabled path-randomization MTD techniques under realistic multi-stage attack scenarios. \sys{} provides a controlled and reproducible enterprise-style environment for jointly evaluating attacker disruption, reconnaissance consistency, network unpredictability, operational overhead, and defender--attacker balance under standardized conditions. Experimental results demonstrate that aggressive path randomization can substantially disrupt multi-stage attack progression while maintaining practical network performance and service stability. More broadly, our findings highlight the importance of deployment-oriented MTD evaluation and show that SDN-based PR-MTD can provide effective security benefits without degrading legitimate network operation.

\bibliographystyle{ACM-Reference-Format}
\bibliography{sample-base}

\vspace{0.5cm}
\section*{Appendix}
\vspace{0.5cm}
\appendix


\section{Ethical Considerations}
\noindent This work was conducted in a fully isolated testbed environment with no connection to public or production networks. All attack scenarios were executed between containerized hosts within MTD-Playground for the sole purpose of evaluating defensive effectiveness. No experiments targeted third-party systems, and no human subjects, personal data, or sensitive information were involved.




\section{Composite Score Analysis}
\label{app:detailed_breakdowns}

\vspace{0.3cm}
\subsection{OCI Ratio Breakdown}
\label{app:oci_breakdown}

Table~\ref{tab:oci_breakdown} shows the individual performance ratios used to compute the Operational Cost Index. These ratios are identical across all scenarios and intervals.

\subsection{Full Interval Metrics}
\label{app:interval_tables}

Table~\ref{tab:interval_full} reports RA, ASR, and ACT at every interval for all three scenarios, computed using Eqs.~\ref{eq:ra}--\ref{eq:act}.

\begin{table}[htbp]
\centering
\caption{OCI ratio breakdown. Each ratio is computed as MTD value / No-MTD value. For throughput, the ratio is inverted so that performance degradation increases cost. OCI is the geometric mean of the eight component ratios.}
\label{tab:oci_breakdown}
\footnotesize
\setlength{\tabcolsep}{5pt}
\renewcommand{\arraystretch}{1.1}
\begin{tabularx}{\columnwidth}{@{}X r@{}}
\hline
\textbf{OCI Ratio} & \textbf{Value} \\
\hline
ONOS CPU & 1.49$\times$ \\
ONOS Memory & 1.47$\times$ \\
Svc latency --- Att$\to$Web & 1.41$\times$ \\
Svc latency --- Web$\to$App & 1.97$\times$ \\
Svc latency --- App$\to$DB & 0.35$\times$ \textit{(improved)} \\
Throughput --- Att$\to$Web & 0.94$\times$ \textit{(improved)} \\
Throughput --- Web$\to$App & 0.97$\times$ \textit{(improved)} \\
Throughput --- App$\to$DB & 0.76$\times$ \textit{(improved)} \\
\hline
\textbf{OCI (geometric mean)} & \textbf{1.049} \\
\hline
\end{tabularx}
\end{table}

\begin{table}[htbp]
\centering
\caption{RA, ASR, and ACT across all scenarios and intervals.}
\label{tab:interval_full}
\footnotesize
\setlength{\tabcolsep}{3pt}
\renewcommand{\arraystretch}{1.15}

\begin{tabular*}{\columnwidth}
{@{\extracolsep{\fill}} l c r r r r @{}}
\hline
\textbf{Metric} & \textbf{Scen.} & \textbf{No-MTD} &
\textbf{MTD10\,s} & \textbf{MTD20\,s} & \textbf{MTD30\,s} \\
\hline
\multirow{3}{*}{RA (\%)} 
& S1 & 100 & 20.0 & 60.0 & 73.3 \\
& S2 & 100 & 20.0 & 60.0 & 73.3 \\
& S3 & 100 & 40.0 & 70.0 & 80.0 \\
\hline
\multirow{3}{*}{ASR (\%)} 
& S1 & 100 & 4.0 & 36.0 & 53.8 \\
& S2 & 100 & 20.0 & 60.0 & 73.3 \\
& S3 & 100 & 16.0 & 49.0 & 64.0 \\
\hline
\multirow{3}{*}{ACT (s)} 
& S1 & 63.2 & 311.1 & 108.7 & 90.3 \\
& S2 & 42.9 & 209.5 & 74.9 & 62.6 \\
& S3 & 63.2 & 160.0 & 94.9 & 84.1 \\
\hline
\end{tabular*}
\end{table}

\vspace{0.5cm}
\section{SDN-Enabled MTD Comparison}
\label{app:mtd_comparison}

Table~\ref{tab:mtd_comparison} provides a detailed comparison of representative SDN-enabled network MTD studies across attack coverage, evaluation metrics, reproducibility, and benchmarking support.

\vspace{0.5cm}
\section{Supplementary Figures}
\begin{figure*}[t]
  \centering
  \includegraphics[width=\textwidth,height=5.5cm,keepaspectratio]{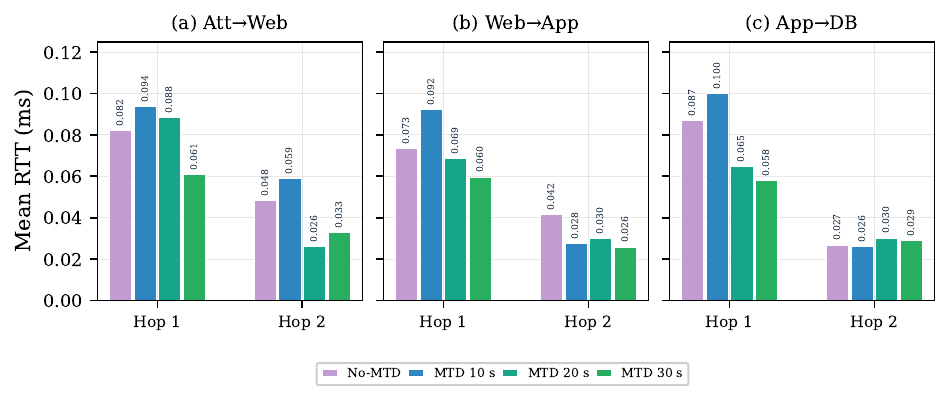}
  \caption{Mean traceroute hop RTT under No-MTD and three MTD intervals. At MTD10\,s, Hop~1 latency exceeds baseline due to frequent flow-rule transitions. At MTD30\,s, Hop~1 drops below baseline as longer-lived rules benefit from OVS fast-path caching.}
  \label{fig:traceroute}
\end{figure*}

\begin{figure*}[t]
  \centering
  \begin{minipage}[t]{0.48\textwidth}
    \centering
    \includegraphics[width=1.0\linewidth]{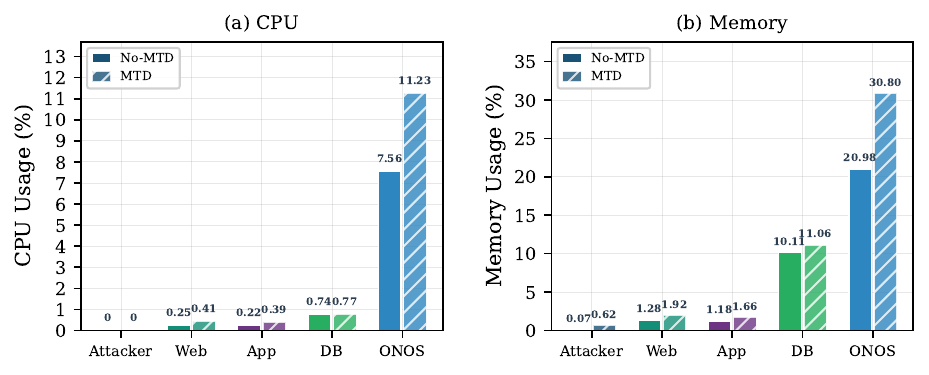}
    \caption{Resource utilisation under No-MTD and MTD. \textit{(a) CPU:} ONOS rises from 7.56\% to 11.23\%; all application containers remain below 1\%. \textit{(b) Memory:} ONOS rises from 20.98\% to 30.80\%.}
    \label{fig:resources}
  \end{minipage}
  \hfill
  \begin{minipage}[t]{0.48\textwidth}
    \centering
    \includegraphics[width=1.1\linewidth]{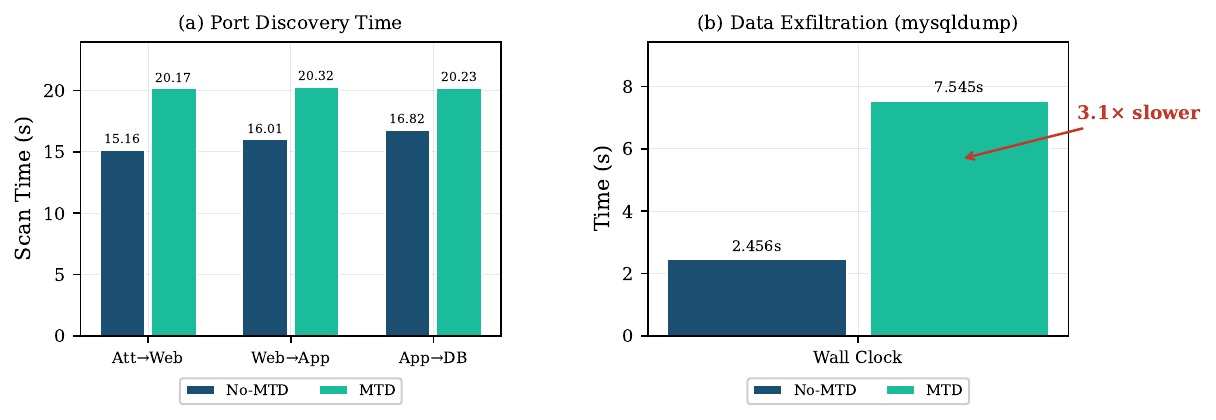}
    \caption{MTD impact on attacker timelines. \textit{(a)} Scan times converge to $\approx$20.2\,s under MTD, eliminating timing-based tier identification. \textit{(b)} Exfiltration slows from 2.456\,s to 7.545\,s (3.1$\times$).}
    \label{fig:scan_exfil}
  \end{minipage}
\end{figure*}

\begin{figure}[t]
  \centering
  \begin{minipage}[t]{0.48\textwidth}
    \centering
    \includegraphics[width=0.7\linewidth]{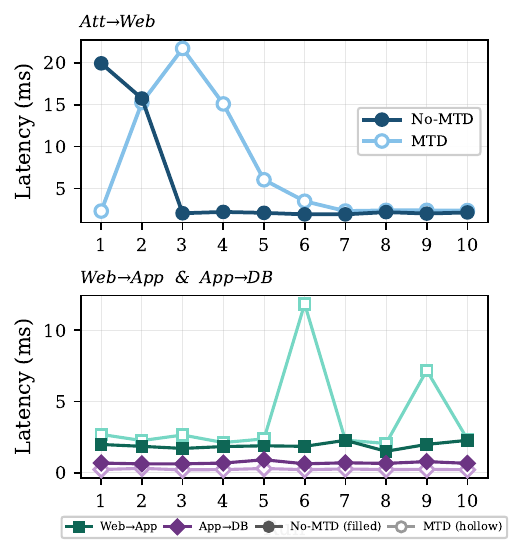}
    \caption{End-to-end service latency (\texttt{curl}) across 10 runs. \textit{Top:} Att$\to$Web latency converges to $\approx$2.3\,ms after initial TCP setup. \textit{Bottom:} App$\to$DB latency drops from 0.7\,ms to 0.24\,ms under MTD due to hop elimination.}
    \label{fig:service_latency}
  \end{minipage}
\end{figure}

\begin{figure}[htbp]
    \centering
    \includegraphics[width=0.67\linewidth]{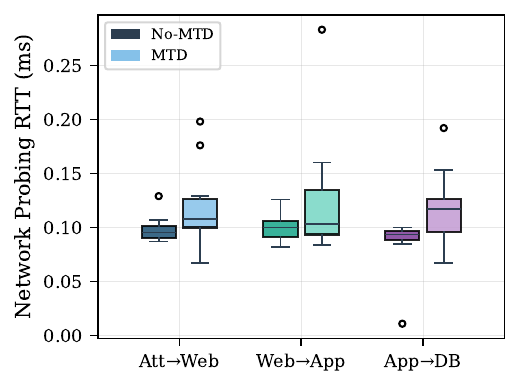}
    \caption{Network probing RTT distribution (ICMP ping, 10 cycles). Under MTD, variance increases by up to 21.3$\times$ on Web$\to$App, making RTT-based network mapping unreliable.}
    \label{fig:probing_rtt}
\end{figure}

This section presents supplementary figures that provide a more detailed view of \sys{}'s network, system, and security behavior. Figures \ref{fig:traceroute} -- \ref{fig:probing_rtt} compare No-MTD and MTD configurations in terms of mean traceroute hop delay, resource utilization, attacker execution time, service latency and RTT variability. Together, these figures support the main findings by illustrating the operational trade-offs and attack-disruption effects of path randomization.
\label{app:figures}

\begin{table*}[t]
\centering
\scriptsize
\caption{Comparison of representative SDN-enabled Network MTD studies across attack coverage, evaluation metrics, reproducibility, and benchmarking support.}
\label{tab:mtd_comparison}
\renewcommand{\arraystretch}{1.2}
\setlength{\tabcolsep}{3pt}
\begin{tabular}{
>{\centering\arraybackslash}p{0.6cm}|
>{\centering\arraybackslash}p{2.5cm}|
>{\centering\arraybackslash}p{1.0cm}|
ccccc|cc|cccc|cc|
>{\centering\arraybackslash}p{2.4cm}|
>{\centering\arraybackslash}p{0.8cm}|
>{\centering\arraybackslash}p{0.8cm}|
>{\centering\arraybackslash}p{0.6cm}
}
\hline
\multirow{3}{*}{\textbf{Study}} 
& \multirow{3}{*}{\makecell{\textbf{What/When/How}\\\textbf{Moves}}}
& \multirow{3}{*}{\makecell{\textbf{Multi-stage}\\\textbf{Attacks}}}
& \multicolumn{13}{c|}{\cellcolor{gray!25}\textbf{Evaluation Metrics}}
& \multirow{3}{*}{\makecell{\textbf{S/P}\\\textbf{Tradeoff}}}
& \multirow{3}{*}{\textbf{Reprod.}}
& \multirow{3}{*}{\makecell{\textbf{Comp./}\\\textbf{Th. Score}}}
& \multirow{3}{*}{\textbf{Ben.}} \\
\cline{4-16}
& & 
& \multicolumn{5}{c|}{\textbf{Security}} 
& \multicolumn{2}{c|}{\textbf{Privacy}} 
& \multicolumn{4}{c|}{\textbf{Funct. / Performance}}
& \multicolumn{2}{c|}{\textbf{Usability}} 
& & & & \\
\cline{4-16}
& & 
& \textbf{ASR} & \textbf{ACT} & \textbf{RA} & \textbf{DT} & \textbf{NU}
& \textbf{DE} & \textbf{FR} 
& \textbf{SL} & \textbf{TH} & \textbf{PLR} & \textbf{SA} 
& \textbf{CO} & \textbf{ED}  
& & & & \\
\hline
\rowcolor{gray!12}
\cite{Jafarian2015} & IP/periodic/SDN flow rules & $\LEFTcircle$ & $\checkmark$ & $\checkmark$ & $\checkmark$ & \ding{55} & \ding{55} & \ding{55} & $\checkmark$ & \ding{55} & \ding{55} & \ding{55} & \ding{55} & $\checkmark$ & Moderate & Security ($\uparrow$) and CO ($\uparrow$) & \ding{55} & \ding{55} & No \\
\hline
\cite{Achleitner2016} & Topo./trig./packet rewrite & \ding{108} & $\checkmark$ & $\checkmark$ & $\checkmark$ & $\checkmark$ & \ding{55} & $\checkmark$ & $\checkmark$ & $\checkmark$ & \ding{55} & \ding{55} & \ding{55} & $\checkmark$ & Moderate & Deception ($\uparrow$), latency ($\uparrow$) & \ding{55} & \ding{55} & No \\
\hline
\rowcolor{gray!12}
\cite{Macwan2019} & IP/periodic/SDN flow rules & $\LEFTcircle$ & $\checkmark$ & \ding{55} & \ding{55} & \ding{55} & \ding{55} & $\checkmark$ & \ding{55} & \ding{55} & \ding{55} & \ding{55} & $\checkmark$ & $\checkmark$ & Easy & \ding{55} & \ding{55} & \ding{55} & No \\
\hline
\cite{Gudla2020} & IP/adaptive/SDN flow rules & $\LEFTcircle$ & $\checkmark$ & \ding{55} & \ding{55} & \ding{55} & \ding{55} & $\checkmark$ & \ding{55} & \ding{55} & \ding{55} & \ding{55} & $\checkmark$ & $\checkmark$ & Moderate & Mutation ($\uparrow$), perf. ($\downarrow$) & \ding{55} & \ding{55} & No \\
\hline
\rowcolor{gray!12}
\cite{Stein2018} & Path/adaptive/BGP routing & $\LEFTcircle$ & $\checkmark$ & \ding{55} & $\checkmark$ & $\checkmark$ & \ding{55} & \ding{55} & \ding{55} & \ding{55} & \ding{55} & \ding{55} & $\checkmark$ & $\checkmark$ & Moderate & Security ($\uparrow$), perf. ($\downarrow$) & \ding{55} & \ding{55} & No \\
\hline
\cite{Narantuya2019} & IP/periodic/SDN flow rules & $\LEFTcircle$ & $\checkmark$ & \ding{55} & $\checkmark$ & \ding{55} & $\checkmark$ & \ding{55} & \ding{55} & \ding{55} & \ding{55} & \ding{55} & $\checkmark$ & $\checkmark$ & Moderate & Security ($\uparrow$) and CO ($\uparrow$) & \ding{55} & \ding{55} & No \\
\hline
\rowcolor{gray!12}
\cite{Hyder2021} & Path/trigger/SDN intent reinstall & $\LEFTcircle$ & $\checkmark$ & \ding{55} & $\checkmark$ & $\checkmark$ & $\checkmark$ & \ding{55} & \ding{55} & $\checkmark$ & \ding{55} & \ding{55} & $\checkmark$ & $\checkmark$ & Moderate & Security ($\uparrow$) and latency ($\uparrow$) & \ding{55} & \ding{55} & No \\
\hline
\cite{Ahmed2016} & VM/periodic/node reincarnation & $\LEFTcircle$ & $\checkmark$ & $\checkmark$ & $\checkmark$ & \ding{55} & \ding{55} & \ding{55} & \ding{55} & $\checkmark$ & \ding{55} & \ding{55} & \ding{55} & $\checkmark$ & Easy & Security ($\uparrow$) and availability ($\downarrow$) & \ding{55} & \ding{55} & No \\
\hline
\rowcolor{gray!12}
\cite{Chowdhary2018} & Ports/trigger/SDN flow rules & \ding{108} & $\checkmark$ & \ding{55} & $\checkmark$ & \ding{55} & \ding{55} & \ding{55} & \ding{55} & \ding{55} & \ding{55} & \ding{55} & $\checkmark$ & $\checkmark$ & Easy & Security ($\uparrow$) and QoS ($\downarrow$) & \ding{55} & $\checkmark$ & Yes \\
\hline
\cite{Chowdhary2017} & Path/adaptive/SDN flow rules & $\LEFTcircle$ & $\checkmark$ & \ding{55} & $\checkmark$ & $\checkmark$ & \ding{55} & \ding{55} & \ding{55} & \ding{55} & $\checkmark$ & \ding{55} & $\checkmark$ & $\checkmark$ & Moderate & Mitigation ($\uparrow$) and SA ($\downarrow$) & \ding{55} & \ding{55} & No \\
\hline
\rowcolor{gray!12}
\cite{Chai2020} & IP/adaptive/vIP proxies & $\LEFTcircle$ & $\checkmark$ & $\checkmark$ & \ding{55} & \ding{55} & \ding{55} & $\checkmark$ & $\checkmark$ & \ding{55} & \ding{55} & $\checkmark$ & $\checkmark$ & $\checkmark$ & Moderate & Security ($\uparrow$) and CO ($\downarrow$) & \ding{55} & \ding{55} & No \\
\hline
\cite{Moghaddam2024} & IP/random/SDN flow rules & $\LEFTcircle$ & \ding{55} & \ding{55} & \ding{55} & $\checkmark$ & \ding{55} & \ding{55} & \ding{55} & \ding{55} & \ding{55} & \ding{55} & \ding{55} & $\checkmark$ & Moderate & Latency ($\downarrow$) and CO ($\uparrow$) & $\checkmark$ ($\LEFTcircle$) & \ding{55} & No \\
\hline
\rowcolor{gray!12}
\cite{Abdelkhalek2022} & Path/periodic/SDN flow rules & $\LEFTcircle$ & $\checkmark$ & \ding{55} & \ding{55} & $\checkmark$ & \ding{55} & \ding{55} & \ding{55} & $\checkmark$ & \ding{55} & $\checkmark$ & $\checkmark$ & $\checkmark$ & Easy & Mitigation ($\uparrow$) and CO ($\downarrow$) & \ding{55} & \ding{55} & No \\
\hline
\cite{Aydeger2025} & Path/trigger/SDN flow rules & $\LEFTcircle$ & $\checkmark$ & \ding{55} & \ding{55} & $\checkmark$ & \ding{55} & \ding{55} & \ding{55} & $\checkmark$ & $\checkmark$ & $\checkmark$ & $\checkmark$ & $\checkmark$ & Moderate & Availability ($\uparrow$) and VNF ($\uparrow$) & \ding{55} & \ding{55} & No \\
\hline
\rowcolor{gray!12}
\cite{Fronczyk2025} & Path/periodic/SDN flow rules & $\LEFTcircle$ & $\checkmark$ & \ding{55} & $\checkmark$ & \ding{55} & \ding{55} & $\checkmark$ & $\checkmark$ & \ding{55} & $\checkmark$ & $\checkmark$ & \ding{55} & $\checkmark$ & Moderate & Security ($\uparrow$) and stability ($\downarrow$) & \ding{55} & \ding{55} & No \\
\hline
\makecell[c]{\textbf{This} \\ \textbf{Work}} & \textbf{Path/periodic/SDN path switching} & \ding{108} & $\checkmark$ & $\checkmark$ & $\checkmark$ & n/a$^{*}$ & $\checkmark$ & $\checkmark$ & \ding{55} & $\checkmark$ & $\checkmark$ & $\checkmark$ & $\checkmark$ & $\checkmark$ & \textbf{Easy} & \textbf{ASR ($\uparrow$), CO ($\uparrow$)} & $\checkmark$ & $\checkmark$ & \textbf{Yes} \\
\hline
\end{tabular}

\scriptsize{

\textbf{ASR} = Attack Success Rate,
\textbf{ACT} = Attack Completion Time,
\textbf{RA} = Reconnaissance Accuracy,
\textbf{DT} = Detection Time,
\textbf{NU} = Network Unpredictability,
\textbf{DE} = Data Exposure Reduced,
\textbf{FR} = Fingerprinting Risk,
\textbf{SL} = Service Latency,
\textbf{TH} = Throughput,
\textbf{PLR} = Packet Loss Rate,
\textbf{SA} = Service Availability,
\textbf{CO} = Controller Overhead,
\textbf{ED} = Ease of Deployment,
\textbf{S/P} = Security vs Performance,
\textbf{Reprod.} = Reproducibility,
\textbf{Comp./Th. Score} = Composite / Threat Score,
\textbf{Ben.} = Benchmarking Environment,
$^{*}$Path mutation is schedule-based, not detection-triggered,
\ding{108} = Full multi-stage,
$\LEFTcircle$ = Partial multi-stage / partial,
$\checkmark$ = Available,
\ding{55} = Not Available.
}
\end{table*}

\clearpage
\end{document}